\documentclass[twocolumn,prb,amsmath,amssymb]{revtex4}
\usepackage[usenames,dvipsnames]{color}

\newcommand \comma {\mbox{\makebox[.1 in]{ },}}
\newcommand \perd {\mbox{\makebox[.1 in]{ }.}}

\def \be {\begin{equation}}
\def \ee {\end{equation}}
\def \ben {\begin{eqnarray}}
\def \een {\end{eqnarray}}

\begin{document}
\bibliographystyle{../prsty}
\draft

\title{Multistep quantum master equation theory for response functions in  four wave mixing electronic spectroscopy of multichromophoric macromolecules}

\author{Seogjoo Jang}
\address{Department of Chemistry and Biochemistry, Queens College of the City University of New York, Flushing, NY 11367} 
\date{Published in the {\it Bulletin of the Korean Chemical Society} {\bf 33}, 997-1008 (2012)}

\begin{abstract}
This work provides an alternative derivation of third order response functions in four wave mixing spectroscopy of  multichromophoric macromolecular systems considering only single exciton states.   For the case of  harmonic oscillator bath linearly and diagonally coupled to exciton states, closed form expressions showing all the explicit time dependences are derived.  These expressions can provide more solid physical basis for understanding 2-dimensional electronic spectroscopy signals.  For more general cases of system-bath coupling, the quantum master equation (QME) approach is employed for the derivation of multistep time evolution equations for Green function-like operators.  Solution of these equations is feasible at the level of 2nd order non-Markovian QME, and the new approach can account for inter-exciton coupling, dephasing, relaxation, and non-Markovian effects in a consistent manner. 
\end{abstract}
\maketitle

\vskip 0.5in
\section{Introduction}
Electronic excitation is the outcome of correlated motion of electrons and is fundamentally quantum mechanical.  When they are put together at nanometer length scale, each excitation loses its individuality and coherent superposition of those excitations, excitons, can be formed.  This is possible  even without physical contacts between chromophores because of long range characteristics of Coulomb interactions in particular.    Thus, the energetics and the dynamics of delocalized excitons in so called multichromophoric macromolecule (MCMM) often result in optical properties that are distinctively different from those of individual chromophores.  Well known examples of such MCMMs are photosynthetic light harvesting complexes,\cite{sundstrom-jpcb103,hu-qrb35,renger-pr343,cogdell-qrb39} conjugated polymers,\cite{heeger-rmp73,gaab-jpcb108,lim-jcp117,beljonne-jpcb109,collini-science323,spano-arpc57} and dendrimers.\cite{kopelman-prl78,hofkens-jacs122,ranasinghe-jacs125,ahn-jpcb110,peng-pr87}  Excitons in these MCMMs are tunable but fragile, which are being utilized positively for efficient and robust collection/transfer of excitons in natural photosynthetic light harvesting complexes.\cite{yang-jppa142,yang-bj85,jang-prl92,jang-jpcb111}  Similar utilization in synthetic MCMMs, if possible, can lead to  novel mechanisms of solar energy conversion\cite{bredas-acr42,chasteen-semsc92} and sensor development. Detailed spectroscopic studies of MCMM are needed to explore these possibilities.

In general, spectroscopic study of  MCMM is difficult because of broad range of dynamical time scales and the large number of structural/energetic degrees of freedom.  In understanding how their optical properties reflect the molecular level structural and dynamical details, conventional linear spectroscopy is severely limited.   Nonlinear spectroscopy of MCMM has an important role to play in this regard.\cite{mukamel,jonas-arpc54,cho-jpcb109,cho-cr108}
Indeed, recent progress in 2-dimensional electronic spectroscopy (2DES)\cite{engel-nature466,collini-nature463,panitchayangkoon-pnas107,schlau-cohen-cp386} made it possible to identify quantum coherence lasting up to ${\rm 500\ fs}$ in photosynthetic light harvesting complexes despite significant amount of disorder and fluctuations.  Theoretical modelings of these 2DES signals have been made,\cite{palmieri-pccp12,sharp-jcp132,abramavicius-jcp133,olbrich-jpcb115,chen-jcp134} but clear understanding of the origins and effects of the coherent quantum beating is not available yet.    This motivates the need for more advanced theoretical approaches that can provide reliable and efficient description of exciton dynamics in such MCMM systems.   The formulation presented in this work provides an important basis for developing such theoretical approaches.

\section{Hamiltonian}
For a molecular system subject to an optical probe,  the total
matter-radiation Hamiltonian (within the semiclassical approximation of the radiation) 
is 
\be
H_T(t)=H+H_{int}(t)\comma
\ee
where $H$ is the material Hamiltonian representing the MCMM and its environment, and $H_{int}(t)$ represents the matter-radiation interaction.  The material Hamiltonian can be expressed as 
\be
H\equiv {\mathcal E}_g|g\rangle\langle g|+H_e+H_{eb}+H_b\comma
\ee
where $|g\rangle$ is the ground electronic state with 
energy ${\mathcal E}_g$, $H_{e}$ the exciton Hamiltonian, $H_{eb}$ the exciton-bath 
Hamiltonian, and $H_b$ the bath Hamiltonian.   Here, the ``bath" represents all other molecular and environmental degrees of freedom interacting with the excitons.  Thus, when the molecule is in the ground electronic state, $H$ reduces to 
\be
H_g={\mathcal E}_g|g\rangle\langle g|+H_b \ . 
\ee
When the molecules are electronically excited, $H$ effectively becomes 
\be
H_{ex}=H_e+H_{eb}+H_b \ .
\ee

For simplicity, only single exciton states are considered here although consideration up to double exciton states may be necessary for full analysis of 2DES signals.  In the site excitation basis,  the exciton Hamiltonian has the following form:
\ben
H_e=\sum_l (E_l+ {\mathcal E}_e) |l\rangle\langle l|+\sum_{l,l'} \Delta_{l,l'} |l\rangle\langle l'|\ ,
\een
where $|l\rangle$ is the state where only the $l$th chromophore is excited and $(E_l+{\mathcal E}_e)$ is its energy.    In this definition, ${\mathcal E}_e$ is a reference excitation energy of the MCMM, which can be chosen to be the average of the excited state energies,  the lowest exciton state, or any other value that can represent the average property of the excitons.   Thus, $E_l$'s are values comparable to differences of excitation energies in the single exciton space.    For later use, we introduce $h_e$ as the exciton Hamiltonian without the contribution of ${\mathcal E}_e$ as follows:
\be
h_e=H_e-{\mathcal E}_e\sum_l|l\rangle\langle l|=\sum_l  E_l|l\rangle\langle l|+\sum_{l,l'} \Delta_{l,l'} |l\rangle\langle l'|\ \ .
\ee

When diagonalized, the exciton Hamiltonian can be expressed as 
\be
H_e=\sum_j ({\mathcal E}_j+{\mathcal E}_e)|\varphi_j\rangle\langle \varphi_j| \comma  \label{eq:he-m1}
\ee 
where ${\mathcal E}_j+{\mathcal E}_e$ is the energy of the exciton state $|\varphi_j\rangle$. Equivalently, this can be expressed as 
\be
h_e=\sum_j {\mathcal E}_j|\varphi\rangle\langle \varphi_j| \ .
\ee
Analogously, we can also define 
\be
h_{ex}=h_e+H_{eb}+H_b \ .
\ee

The unitary transformation matrix between the site basis and the exciton basis is denoted as $U$.  The matrix elements of this transformation are defined as $U_{lj}=\langle l|\varphi_j\rangle$, and the following relation holds: 
\be
|l\rangle =\sum_{j=1}^N U_{lj}^* |\varphi_j\rangle\ .
\ee

The transition dipole vector for the excitation from $|g\rangle$ to $|l\rangle$ is denoted as $\mbox{\boldmath$\mu$}_l$.  Then, the total electronic polarization operator for the transitions to the single exciton space of the MCMM is given by 
\ben
{\bf P}&=&\sum_{l} \mbox{\boldmath$\mu$}_l \left (|l\rangle \langle g|+|g\rangle\langle l|\right)  \nonumber \\
&=&\sum_l\sum_j \left (\mbox{\boldmath$\mu$}_l U_{lj}^* |\varphi_j\rangle\langle g|+|g\rangle \langle \varphi_j| U_{lj} \mbox{\boldmath$\mu$}_l \right ) \nonumber \\
&=&\sum_j \left ({\bf D}_j |\varphi_j\rangle\langle g|+|g\rangle\langle \varphi_j| {\bf D}_j  \right)  \nonumber \\
&=&|{\bf D}\rangle\langle g|+|g\rangle\langle {\bf D}|\ , \label{eq:p_def}
\een
where ${\bf D}_j=\sum_l \mbox{\boldmath$\mu$}_l U_{lj}^*$ and $|{\bf D}\rangle\equiv \sum_j {\bf D}_j |\varphi_j\rangle$. 

Assuming three incoming pulses, the  matter-radiation interaction Hamiltonian is   
\ben
&&H_{int}(t)=\sum_{\alpha=1}^3 {\bf E}_\alpha (t-t_\alpha)\cdot| {\bf D} \rangle \langle g|e^{i{\bf k}_\alpha\cdot{\bf r}-i\omega_\alpha t}+{\rm H.c.} \nonumber \\
&&=\sum_{\alpha=1}^3\sum_j {\bf E}_\alpha(t-t_\alpha)\cdot  {\bf D}_j |\varphi_j\rangle \langle g| e^{i{\bf k}_\alpha\cdot {\bf r}-i\omega_\alpha t} + {\rm H.c.}  \nonumber \\
&&=\sum_{\alpha=1}^3 E_\alpha(t-t_\alpha) |D_\alpha \rangle \langle g| e^{i{\bf k}_\alpha\cdot {\bf r}-i\omega_\alpha t} + {\rm H.c.} \ ,\label{eq:h_int_def}
\een    
where $E_\alpha (t-t_\alpha)$ is the amplitude of the $\alpha$th pulse with polarization vector $\mbox{\boldmath{${\epsilon}$}}_\alpha$. Thus,  ${\bf E}_\alpha (t-t_\alpha)=E_\alpha (t-t_\alpha) \mbox{\boldmath{${\epsilon}$}}_\alpha $.   It is assumed that $t_3 \geq t_2\geq t_1$. 
In the last line of Eq. (\ref{eq:h_int_def}), $|D_\alpha\rangle$ is the sum of all the exciton states weighted by the components of the transition dipoles along the direction $\mbox{\boldmath{${\epsilon}$}}_\alpha$ and has the following expression:
\be 
|D_\alpha\rangle=\sum_j \mbox{\boldmath{${\epsilon}$}}_\alpha \cdot  {\bf D}_j |\varphi_j\rangle \ .
\ee  

\section{Response functions}
\subsection{General expressions}
Assume that the optical field is active from $t=0$ and that the total density operator
of the material at this time is $\rho (0)=|g\rangle \langle g| \rho_b $, where $\rho_b=e^{-\beta H_b}/Tr_b\{e^{-\beta H_b}\}$ with $\beta=1/k_BT$.  Thus, $[\rho_b,H_b]=0$. Then, the time evolution operator governing the total material system for $t>0$ is given by 
\be
{\mathcal U}(t)=\exp_{(+)}\left\{-\frac{i}{\hbar}\int_0^t dt' H_T(t')\right\}\ ,
\ee
where $(+)$ represents chronological time ordering.
Then, the total density operator at time $t_m$ is 
\be
\rho(t_m)={\mathcal U}(t_m)\rho(0) {\mathcal U}^\dagger(t_m) \ .
\ee
Expanding ${\mathcal U}(t)$ and ${\mathcal U}^\dagger(t)$ with respect to $H_{int}(t)$, and collecting all the terms of the third order, we find the following third order components of the density operator: 
\be
\rho^{(3)}(t_m)=\rho_{I}(t_m)+\rho_{I}^\dagger(t_m)+\rho_{II}(t_m)+\rho_{II}^\dagger (t_m) \comma
\ee
where
\ben
\rho_{I}(t_m)&=&-\frac{i}{\hbar^3}\int_{0}^{t_m} dt\int_{0}^{t} dt' \int_{0}^{t_m} dt'' e^{-iH(t_m-t)/\hbar} \nonumber \\
&&\times H_{int}(t)e^{-iH(t-t')/\hbar}H_{int}(t')e^{-iHt'/\hbar}\rho(0)\nonumber \\
&&\times e^{iHt''/\hbar}H_{int}(t'')e^{iH(t_m-t'')/\hbar} \ , \nonumber\\ \label{eq:rho_I}
\\
\rho_{II}(t_m)&=&-\frac{i}{\hbar^3}\int_{0}^{t_m} dt \int_{0}^{t} dt' \int_{0}^{t'} dt'' e^{-iHt_m/\hbar} \rho (0) \nonumber \\
&&\times e^{iHt''/\hbar} H_{int}(t'') e^{iH(t'-t'')/\hbar} H_{int}(t') \nonumber \\
&&\times e^{iH(t-t')/\hbar} H_{int}(t)e^{iH(t_m-t)/\hbar} \ .  \nonumber\\ \label{eq:rho_II}
\een
In Eq. (\ref{eq:rho_I}),
the integration over $t''$ can be split into three regions,
$0 < t''<t'$, $t'<t''<t$, and $t <t''<t_m$.    Relabeling the dummy time integration variables in each region such that $t\geq t' \geq t''$,  the three terms can be rewritten so as to have the same time integration boundaries as $\rho_{II}(t_m)$.  The resulting third order components can be expressed as 
\ben
\rho^{(3)}(t_m)&=&-\frac{i}{\hbar^3}\int_{0}^{t_m} dt \int_{0}^t dt' \int_{0}^{t'} dt''\sum_{j=1}^4 {\mathcal T}_j (t_m,t,t',t'') \nonumber \\
&&+{\rm H.c.}\ , \label{eq:rho_3}
\een 
where
\ben
&&{\mathcal T}_1(t_m,t,t',t'')\equiv  e^{-iH(t_m-t')/\hbar}H_{int}(t')e^{-iH(t'-t'')/\hbar}\nonumber \\
&&\hspace{.2in}\times H_{int}(t'') e^{-iHt''/\hbar}\rho(0)e^{iHt/\hbar}H_{int}(t) e^{iH(t_m-t)/\hbar} \label{eq:t1}\ , \nonumber \\ \\
&&{\mathcal T}_2(t_m,t,t',t'')\equiv e^{-iH(t_m-t)/\hbar}H_{int}(t)e^{-iH(t-t'')/\hbar}\nonumber \\
&&\hspace{.2in} \times H_{int}(t'') e^{-iHt''/\hbar}\rho(0)e^{iHt'/\hbar}H_{int}(t') e^{iH(t_m-t')/\hbar} \ ,\nonumber \\ \label{eq:t2}\\
&&{\mathcal T}_3(t_m,t,t',t'')\equiv e^{-iH(t_m-t)/\hbar}H_{int}(t)e^{-iH(t-t')/\hbar}\nonumber \\
&&\hspace{.2in}\times H_{int}(t')e^{-iHt'/\hbar}\rho(0)e^{iHt''/\hbar}H_{int}(t'') e^{iH(t_m-t'')/\hbar} \ ,\nonumber \\ \label{eq:t3}\\
&&{\mathcal T}_4(t_m,t,t',t'')\equiv  e^{-iHt_m/\hbar}\rho(0)e^{iHt''/\hbar}H_{int}(t'')\nonumber \\
&&\hspace{.2in} \times e^{iH(t'-t'')/\hbar} H_{int}(t')e^{iH(t-t')/\hbar}H_{int}(t)e^{iH(t_m-t)/\hbar} \ . \nonumber \\ \label{eq:t4}
\een
In the above expressions, Eqs. (\ref{eq:t1})-(\ref{eq:t3}) come from $\rho_{I}(t_m)$ and Eq. (\ref{eq:t4}) comes from  $\rho_{II}(t_m)$. 

The corresponding third order contribution to the polarization can be calculated by taking the trace of the scalar product between ${\bf P}$, Eq. (\ref{eq:p_def}), and $\rho^{(3)}(t_m)$, Eq. (\ref{eq:rho_3}).  The resulting expression for the third order polarization at time $t_m$  can be shown to be
\ben
&&\bar{\bf P}^{(3)}(t_m)\equiv Tr\{{\bf P}\rho^{(3)}(t_m)\} \nonumber \\
&&=\frac{2}{\hbar^3}\sum_{j=1}^4 {\rm Im}\int_{0}^{t_m}dt\int_{0}^{t} dt' \int_{0}^{t'}dt''\ Tr\left\{{\bf P} {\mathcal T}_j (t_m,t,t',t'')\right\} \ , \nonumber \\ \label{eq:p4t}
\een
where $``{\rm Im}"$ implies imaginary part of the complex function.

Denote the unit vector of the polarization being measured at time $t_m$ as $\mbox{\boldmath{${\epsilon}$}}_m$.  Taking scalar product of this with the integrand of Eq. (\ref{eq:p4t}) and considering only those terms where interactions with $E_1$, $E_2$, and $E_3$  occur in the chronological order at $t''$, $t'$, and $t$, respectively, we obtain the following general expression: 
\ben
&&\mbox{\boldmath{${\epsilon}$}}_m\cdot Tr\left\{{\bf P} {\mathcal T}_1(t_m,t,t',t'')\right\}= E^*_3 (t-t_3)E^*_2 (t'-t_2)\nonumber \\
&&\hspace{.3in}\times E_1 (t''-t_1)e^{i(\omega_3t+\omega_2t'-\omega_1 t'')}e^{-i({\bf k}_3+{\bf k}_2 -{\bf k}_1)\cdot {\bf r}} \nonumber \\
&&\hspace{.3in}\times Tr_b\left\{ e^{-i H_g (t_m-t')/\hbar}\langle D_2| e^{-iH_{ex} (t'-t'')/\hbar}|D_1\rangle \right . \nonumber \\
&&\hspace{.4in}\times \left . \rho_b e^{iH_g (t-t'')/\hbar}\langle D_3|  e^{iH_{ex}(t_m-t)/\hbar}|D_4\rangle\right\} \ , \nonumber \\ \label{eq:pt1}  
\een
\ben
&&\mbox{\boldmath{${\epsilon}$}}_m\cdot Tr\left\{ {\bf P} {\mathcal T}_2(t_m,t,t',t'')\right\}=E^*_3 (t-t_3)E^*_2 (t'-t_2)\nonumber \\
&&\hspace{.3in} \times E_1 (t''-t_1)e^{i(\omega_3 t+\omega_2 t'-\omega_1 t'')} e^{-i({\bf k}_3 +{\bf k}_2 -{\bf k}_1)\cdot{\bf r}}\nonumber \\
&&\hspace{.3in}\times Tr_b\left\{e^{-iH_g(t_m-t)/\hbar}\langle D_3| e^{-iH_{ex} (t-t'')/\hbar}|D_1\rangle \right . \nonumber \\
&&\hspace{.4in} \times \left . \rho_b e^{iH_g (t'-t'')/\hbar} \langle D_2| e^{iH_{ex}(t_m-t')/\hbar}|D_4\rangle \right\} \ , \nonumber \\ \label{eq:pt2}
\een
\ben
&&\mbox{\boldmath{${\epsilon}$}}_m\cdot Tr\left\{{\bf P} {\mathcal T}_3(t_m,t,t',t'')\right\}=E^*_3(t-t_3) E_2 (t'-t_2)\nonumber \\
&&\hspace{.3in} \times E^*_1(t''-t_1)e^{i(\omega_3 t-\omega_2t'+\omega_1t'')} e^{-i({\bf k}_3-{\bf k}_2+{\bf k}_1)\cdot {\bf r}} \nonumber \\
&&\hspace{.3in}\times Tr_b\left\{e^{-iH_g(t_m-t)/\hbar}\langle D_3| e^{-iH_{ex} (t-t')/\hbar}|D_2\rangle \right . \nonumber \\
&&\hspace{.4in} \times \left . e^{-iH_g(t'-t'')/\hbar} \rho_b \langle D_1| e^{iH_{ex}(t_m-t'')/\hbar} |D_4\rangle \right\} \ , \nonumber \\ \label{eq:pt3} 
\een
\ben
&&\mbox{\boldmath{${\epsilon}$}}_m\cdot Tr\left\{ {\bf P} {\mathcal T}_4(t_m,t,t',t'')\right\}= E_3^*(t-t_3)E_2(t'-t_2)\nonumber \\
&&\hspace{.3in} \times E_1^* (t''-t_1) e^{i(\omega_3t-\omega_2t'+\omega_1 t'')} e^{-i({\bf k}_3-{\bf k}_2+{\bf k}_1)\cdot {\bf r}} \nonumber  \\
&&\hspace{.3in} \times Tr_b\left\{e^{-iH_g (t_m-t'')/\hbar}\rho_b \langle D_1|e^{i H_{ex} (t'-t'')/\hbar}|D_2\rangle \right . \nonumber \\
&&\hspace{.4in} \left . \times e^{iH_g(t-t')/\hbar} \langle D_3| e^{iH_{ex} (t_m-t)/\hbar}|D_4\rangle \right\} \  . \nonumber \\  \label{eq:pt4}
\een
where
\be
|D_4\rangle=\sum_j \mbox{\boldmath{${\epsilon}$}}_m \cdot {\bf D}_j |\varphi_j\rangle \ .
\ee
At this point, it is convenient to introduce new time variables,  $\tau=t_m-t$, $T_p=t-t'$, and $\tau'=t'-t''$. As will be clear, these notations imply that $\tau$ and $\tau'$ are coherence times and that $T_p$ is the population time.  Replacing the time integration variables in Eq. (\ref{eq:p4t}) with these and taking  scalar product of $\bar {\bf P}^{(3)}(t_m)$ with $\mbox{\boldmath{$\epsilon$}}_m$, we obtain the following expression:

\ben
&&\mbox{\boldmath{${\epsilon}$}}_m\cdot\bar{\bf P}^{(3)}(t_m) \nonumber \\
&&=\frac{2}{\hbar^3}\sum_{j=1}^4 {\rm Im}\int_{0}^{t_m}d\tau \int_0^{t_m-\tau} d T_p \int_{0}^{t_m-T_p-\tau} d \tau' \nonumber \\
&& \mbox{\boldmath{${\epsilon}$}}_m\cdot Tr\left\{{\bf P} {\mathcal T}_j (t_m,t_m-\tau,t_m-\tau-T_p,t_m-\tau-T_p-\tau')\right\} \nonumber \\ \label{eq:p4t-2}
\een

Inserting Eqs. (\ref{eq:pt1})-(\ref{eq:pt4}) into Eq. (\ref{eq:p4t-2}), 
\ben
&&\mbox{\boldmath{${\epsilon}$}}_m \cdot \bar{{\bf P}}^{(3)}(t_m)=2\ {\rm Im}\ \int_{0}^{t_m}d\tau\int_0^{t_m-\tau} dT_p \int_{0}^{t_m-T_p-\tau} d\tau' \nonumber \\
&&\hspace{.1in}  \left\{ E_3^*(t_m-\tau-t_3)E_2^*(t_m-\tau-T_p-t_2)\nonumber \right .\\
&&\hspace{.1in} \times E_1(t_m-\tau-T_p-\tau'-t_1)e^{i(\omega_3+\omega_2-\omega_1)(t_m-\tau)}\nonumber \\
&&\hspace{.1in}\times e^{-i(\omega_2-\omega_1)T_p}e^{i\omega_1 \tau'} e^{i({\mathcal E}_e-{\mathcal E}_g)(\tau-\tau')/\hbar}\nonumber \\
&&\hspace{.1in}\times e^{-i({\bf k}_3+{\bf k}_2- {\bf k}_1)\cdot {\bf r}}\left ( \chi^{(1)}(\tau,T_p,\tau')+ \chi^{(2)}(\tau,T_p,\tau')\right )\nonumber \\
&&+E_3^* (t_m-\tau-t_3) E_2(t_m-\tau-T_p-t_2)\nonumber \\
&&\hspace{.1in} \times E_1^*(t_m-\tau-T_p-\tau'-t_1) e^{i(\omega_3-\omega_2+\omega_1)(t_m-\tau)} \nonumber \\
&&\hspace{.1in}\times e^{i(\omega_2-\omega_1)T_p}e^{-i\omega_1 \tau'} e^{i({\mathcal E}_e-{\mathcal E}_g )(\tau+\tau')/\hbar}\nonumber \\
&&\hspace{.1in}\times  e^{-i({\bf k}_3-{\bf k}_2+{\bf k}_1)\cdot {\bf r}}\left .\left (\chi^{(3)}(\tau,T_p,\tau')+\chi^{(4)}(\tau,T_p,\tau')\right ) \right\}\ , \nonumber \\
\een
where
\ben
&&\chi^{(1)}(\tau,T_p,\tau')=Tr_b\left\{ e^{-i H_b (\tau+T_p)/\hbar}\langle D_2| e^{-ih_{ex} \tau'/\hbar}|D_1\rangle \right . \nonumber \\
&&\hspace{.1in} \left . \times \rho_b e^{iH_b(T_p+\tau')/\hbar} \langle D_3| e^{ih_{ex} \tau/\hbar} |D_4\rangle \right\}\ ,\label{eq:resp1}
\een
\ben
&&\chi^{(2)}(\tau,T_p,\tau')= Tr_b\left\{ e^{-i H_b \tau/\hbar}\langle D_3| e^{-ih_{ex} (T_p +\tau')/\hbar}|D_1\rangle  \right . \nonumber \\
&&\hspace{.1in}\left . \times \rho_b e^{iH_b \tau'/\hbar} \langle D_2| e^{ih_{ex}(\tau+T_p)/\hbar} |D_4\rangle \right\}\ ,\label{eq:resp2}
\een
\ben
&&\chi^{(3)}(\tau,T_p,\tau')= Tr_b\left\{ e^{-i H_b \tau/\hbar }\langle D_3| e^{-ih_{ex} T_p/\hbar }|D_2\rangle   \nonumber \right .\\
&&\hspace{.1in} \left . \times e^{-iH_b \tau'/\hbar}\rho_b\langle D_1| e^{ih_{ex}(\tau+T_p+\tau')/\hbar} |D_4\rangle \right\}\ ,\hspace{.1in}\label{eq:resp3}
\een
\ben
&&\chi^{(4)}(\tau,T_p,\tau')= Tr_b\left\{e^{-iH_b(\tau+T_p+\tau')/\hbar}\rho_b \right . \nonumber \\
&&\hspace{.1in}\left . \times \langle D_1|e^{i h_{ex} \tau'/\hbar }|D_2\rangle e^{iH_b T_p/\hbar} \langle D_3| e^{ih_{ex} \tau/\hbar }|D_4 \rangle  \right\} \ . \hspace{.2in}\label{eq:resp4}
\een
The above general expressions for the response functions are equivalent to those in previous works\cite{mukamel,jonas-arpc54,cho-jpcb109,cho-cr108,schlau-cohen-cp386} within the assumption that only single exciton states contribute.  Fourier transforms of these with respect to $\tau$ and $\tau'$  can be related to the spectra of 2DES if proper averaging over the ensemble of disorder is made.

\subsection{Closed form expressions for diagonal system-bath coupling in the exciton basis}

For the case where the bath Hamiltonian can be modeled by harmonic oscillators and the exciton-bath coupling is diagonal in the exciton basis and is linear in the displacements of harmonic oscillators, a simple 
closed-form expression can be found for each response function.   
Thus, suppose that $H_b=\sum_n\hbar \omega_n(b_n^\dagger b_n+1/2)$, where where $b_n$ and $b_n^\dagger$ are the lowering and raising operators
of the $n$th oscillator, and that 
\ben
&&H_{eb}=\sum_j \delta H_{bj}|\varphi_j\rangle\langle \varphi_j|\nonumber \\
&&=\sum_j\sum_n \hbar \omega_n g_{j,n}(b_n+b_n^\dagger)|\varphi_j\rangle\langle \varphi_j| \comma \label{eq:heb-m1}
\een  
Then, the four response functions defined by Eqs. (\ref{eq:resp1})-(\ref{eq:resp4}) can be calculated explicitly.  Explicit expressions for these can be derived based on generalized cumulant approach.\cite{sung-jcp115}  Alternatively, polaron displacement operator can be used as described in Appendix A. The resulting expressions are as follows.   
\begin{widetext}
\ben
\chi^{(1)}(\tau,T_p,\tau')&=&\sum_j\sum_{j'} D_{2,j}^*D_{1,j}D_{j',3}^*D_{j',4}e^{i\tilde {\mathcal E}_{j'}\tau/\hbar-i\tilde {\mathcal E}_j\tau'/\hbar}\nonumber \\
&&\times e^{-\sum_n g_{j,n}^2\left (\coth(\frac{\beta\hbar\omega_n}{2})(1-\cos(\omega_n\tau'))+i\sin(\omega_n\tau')\right )}\nonumber \\
&&\times e^{-\sum_n g_{j',n}^2\left (\coth(\frac{\beta\hbar\omega_n}{2})(1-\cos(\omega_n\tau))-i\sin(\omega_n\tau)\right )} \nonumber\\
&&\times e^{-\sum_n g_{j,n}g_{j',n}\coth(\frac{\beta\hbar\omega_n}{2}) \{ \cos (\omega_n T_p)-\cos (\omega_n (\tau+T_p))-\cos (\omega_n (T_p+\tau'))+\cos (\omega_n (\tau+T_p+\tau'))\}} \nonumber \\
&&\times e^{i\sum_n g_{j,n} g_{j,n'} \{\sin (\omega_n T_p)-\sin (\omega_n (\tau+T_p))-\sin (\omega_n (T_p+\tau'))+\sin (\omega_n (\tau+T_p+\tau'))\} } \ ,\label{eq:chi1-m1-f} 
\een
\ben
\chi^{(2)}(\tau,T_p,\tau')&=&\sum_j\sum_{j'} D_{3,j}^*D_{1,j}D_{2,j'}^*D_{4,j'}e^{i\tilde {\mathcal E}_{j'}\tau/\hbar-i(\tilde {\mathcal E}_j -\tilde {\mathcal E}_{j'})T_p/\hbar -i\tilde {\mathcal E}_j \tau'/\hbar }\nonumber \\
&&\times e^{-\sum_n g_{j,n}^2\left (\coth(\frac{\beta\hbar\omega_n}{2})(1-\cos(\omega_n(T_p+\tau')))-i\sin(\omega_n(T_p+\tau'))\right )}\nonumber \\
&&\times e^{-\sum_n g_{j',n}^2\left (\coth(\frac{\beta\omega_n}{2})(1-\cos(\omega_n(\tau+T_p))+i\sin(\omega_n(\tau+T_p))\right )} \nonumber\\
&&\times e^{-\sum_n g_{j,n}g_{j',n}\coth(\frac{\beta\hbar\omega_n}{2}) \{ \cos (\omega_n T_p)-\cos (\omega_n \tau)-\cos (\omega_n \tau')+\cos (\omega_n (\tau+T_p+\tau'))\}} \nonumber \\
&&\times e^{i\sum_n g_{j,n} g_{j,n'} \{-\sin (\omega_n T_p)-\sin (\omega_n \tau)-\sin (\omega_n \tau')+\sin (\omega_n (\tau+T_p+\tau'))\} } \ ,  \label{eq:chi2-m1-f} 
\een
\ben  
\chi^{(3)}(\tau,T_p,\tau')&=&\sum_j\sum_{j'} D_{3,j}^*D_{2,j}D_{1,j'}^*D_{4,j'}e^{i\tilde {\mathcal E}_{j'}(\tau+\tau')/\hbar-i(\tilde {\mathcal E}_j-\tilde{\mathcal E}_{j'})T_p/\hbar}\nonumber \\
&&\times e^{-\sum_n g_{j,n}^2\left (\coth(\frac{\beta\hbar\omega_n}{2})(1-\cos(\omega_nT_p))+i\sin(\omega_nT_p)\right )}\nonumber \\
&&\times e^{-\sum_n g_{j',n}^2\left (\coth(\frac{\beta\hbar\omega_n}{2})(1-\cos(\omega_n(\tau+T_p+\tau'))-i\sin(\omega_n(\tau+T_p+\tau'))\right )} \nonumber\\
&&\times e^{-\sum_n g_{j,n}g_{j',n}\coth(\frac{\beta\hbar\omega_n}{2}) \{ \cos (\omega_n (T_p+\tau'))-\cos (\omega_n \tau)-\cos (\omega_n \tau')+\cos (\omega_n (\tau+T_p))\}} \nonumber \\
&&\times e^{i\sum_n g_{j,n} g_{j,n'} \{-\sin (\omega_n (T_p+\tau'))-\sin (\omega_n \tau)+\sin (\omega_n \tau')+\sin (\omega_n (\tau+T_p))\} } \ ,  \label{eq:chi3-m1-f} 
\een
\ben  
\chi^{(4)}(\tau,T_p,\tau')&=&\sum_j\sum_{j'} D_{1,j}^*D_{2,j}D_{3,j'}^*D_{4,j'}e^{i\tilde {\mathcal E}_j\tau'/\hbar+i\tilde {\mathcal E}_{j'}\tau/\hbar}\nonumber \\
&&\times e^{-\sum_n g_{j,n}^2(\coth(\frac{\beta\hbar\omega_j}{2})(1-\cos(\omega_n\tau'))-i\sin(\omega_n\tau'))}\nonumber \\
&&\times e^{-\sum_n g_{j',n}^2(\coth(\frac{\beta\hbar\omega_n}{2})(1-\cos(\omega_n\tau))-i\sin(\omega_n\tau))} \nonumber\\
&&\times e^{-\sum_n g_{j,n}g_{j',n}\coth(\frac{\beta\hbar\omega_n}{2}) \{ \cos (\omega_n (T_p+\tau'))-\cos (\omega_n T_p)-\cos (\omega_n (\tau+T_p+\tau'))+\cos (\omega_n (\tau+T_p))\}} \nonumber \\
&&\times e^{i\sum_n g_{j,n} g_{j,n'} \{-\sin (\omega_n (T_p+\tau'))+\sin (\omega_n T_p)+\sin (\omega_n (\tau+T_p+\tau'))-\sin (\omega_n (\tau+T_p))\} } \ . 
 \label{eq:chi4-m1-f}
\een

\end{widetext}
In the above expressions, definitions of $D_{\alpha,j}$ and $\tilde {\mathcal E}_j$ can be found in Eqs. (\ref{eq:dal_j}) and (\ref{eq:til_ej}).

Despite the simplicity of underlying assumption, the expressions shown in Eqs. (\ref{eq:chi1-m1-f})-(\ref{eq:chi4-m1-f}) are quite complicated, which provide a glimpse of possible complications in reliable modeling of 2DES signals.  First of all, it is obvious that the response functions cannot be simplified into products of linear-spectroscopic lineshape functions as long as there are  common bath modes coupled to different exciton states.     Considering the delocalized nature of excitons, it would be extremely rare to find the case where bath modes coupled to different exciton states are independent of each other. 
Second, terms oscillatory with respect to $T_p$ can be found even in the absence of inter-exciton systen-bath coupling.  This is  because transition to different exciton states is possible during the interaction with the pulse.   This way, coherence between different exciton states can be maintained even in the absence of  bath-mediated inter-exciton couplings.  

\section{Quantum Master Equation for general system-bath coupling}

Let us consider the general situation where the system-bath coupling and 
the bath Hamiltonian can be arbitrary.  The quantum master 
equation (QME) approach\cite{jang-jcp116,jang-jcp118-1} can be employed for the calculation of 
response functions in this case.   First, using the cyclic symmetry within the trace operation and the fact that the exciton states commute with the bath Hamiltonian, Eqs. (\ref{eq:resp1})-(\ref{eq:resp4}) can be recast into forms amenable for QME approach as follows:
\ben
\chi^{(1)}(\tau,T_p,\tau')&&=Tr_{b,e}\left\{e^{-iH_b(\tau+T_p)/\hbar}|D_4\rangle\langle D_3| e^{-ih_{ex}\tau'/\hbar}\nonumber \right .\\ 
&&\times \left . \rho_b|D_2\rangle\langle D_1|e^{iH_b(\tau'+T_p)/\hbar}e^{ih_{ex}\tau/\hbar}\right\}  \ , \\
\chi^{(2)}(\tau,T_p,\tau')&&=Tr_{b,e}\left\{e^{-iH_b\tau/\hbar}|D_4\rangle\langle D_3|e^{-ih_{ex}(T_p+\tau')/\hbar}\nonumber  \right . \\&&\left . \times \rho_b|D_1\rangle\langle D_2| e^{iH_b\tau'/\hbar}e^{ih_{ex}(\tau+T_p)/\hbar}\right\}\ , \\
\chi^{(3)}(\tau,T_p,\tau')&&=Tr_{b,e}\left\{e^{-iH_b\tau/\hbar}|D_4\rangle\langle D_3|e^{-ih_{ex}T_p/\hbar}e^{-iH_b\tau'/\hbar}\right . \nonumber \\
&& \times \left. \rho_b|D_2\rangle\langle D_1|e^{ih_{ex}(\tau+T_p+\tau')/\hbar}\right\} \ , \\
\chi^{(4)}(\tau,T_p,\tau')&&=Tr_{b,e}\left\{e^{-iH_b(\tau+T_p+\tau')/\hbar}\rho_b|D_4\rangle \langle D_1| \right .\nonumber \\
&& \times \left .e^{ih_{ex}\tau'/\hbar}e^{iH_bT_p/\hbar}|D_2\rangle\langle D_3|e^{ih_{ex}\tau/\hbar}\right\}\ .
\een
Detailed methods to calculate these response functions based on the QME approach are described below.

\subsection{Calculation of $\chi^{(1)}(\tau,T_p,\tau')$}

For the calculation of $\chi^{(1)}(\tau,T_p,\tau')$,
let us define
\ben
&&G_1^{(1)}(\tau')=e^{-ih_{ex}\tau'/\hbar}\rho_b|D_2\rangle\langle D_1|e^{iH_b\tau'/\hbar} \ ,\\
&&G_2^{(1)}(T_p,\tau')=e^{-iH_bT_p/\hbar}G_1^{(1)}(\tau')e^{iH_bT_p/\hbar}\ ,\\
&&G_3^{(1)}(\tau,T_p,\tau')=e^{-iH_b\tau/\hbar}|D_4\rangle\langle D_3|G_2^{(1)}(T_p,\tau')e^{ih_{ex}\tau/\hbar} \ . \nonumber \\
\een
Define $g_1^{(1)}(\tau')$, $g_2^{(1)}(T_p,\tau')$, and $g_3^{(1)}(\tau,T_p,\tau')$ as traces of $G_1^{(1)}(\tau')$, $G_2^{(1)}(T_p,\tau')$, and $G_3^{(1)}(\tau,T_p,\tau')$ over the bath degrees of freedom, respectively.
Then, 
\be 
\chi^{(1)}_3(\tau,T_p,\tau')=Tr_e\{g_3^{(1)}(\tau,T_p,\tau')\} \ .
\ee

Unlike the conventional  QME, three time arguments are involved in $g_3^{(1)}(\tau,T_p,\tau')$.  The time evolution with respect to $\tau$ can be considered first.   Appendix B provides a detailed description. The resulting QME has an inhomogeneous term, which in turn depends on the prior time evolutions during $T_p$ and $\tau'$.  Different QMEs have to be derived for these as well.   As can be seen in the Appendix  B.1,  explicit time evolution for $T_p$ is not necessary for the present case.  However, explicit time evolution with respect to $\tau'$ is needed.  Appendix B.1 provides formally exact QMEs for the following two reduced system operators: 
\ben
&&\tilde g_1^{(1)}(\tau')=e^{ih_e\tau'/\hbar} g_1(\tau')\ ,\\
&&\tilde g_3^{(1)}(\tau,T_p,\tau')=g_3^{(1)}(\tau,T_p,\tau') e^{-ih_e\tau'/\hbar}\ .
\een 
Within the 2nd order approximation, all the time evolution operators involving ${\mathcal Q}_L \tilde H_{eb}$ or $\tilde H_{eb} {\mathcal Q}_R$ in Eqs. (\ref{eq:g1p}) and (\ref{eq:g3p-1-final}) can be disregarded.  Thus, we obtain the following approximations: 
\ben
&&\frac{d}{d\tau'}\tilde g_1^{(1)}(\tau')\approx \nonumber \\
&&-\frac{1}{\hbar^2}\int_0^{\tau'} ds\  Tr_b\left\{\rho_b \tilde H_{eb}(\tau' )\tilde H_{eb}(s)\right\} \tilde g_1^{(1)}(s) \ , \\
 &&\frac{d}{d\tau}\tilde g_3^{(1)}(\tau,T_p,\tau')=- \frac{1}{\hbar^2}|D_4\rangle\langle D_3|  e^{-i h_e \tau'/\hbar}   \nonumber \\
&&\hspace{.3in} \times \int_0^{\tau'} ds Tr_b\left\{\tilde H_{eb}(s)\rho_b \tilde g_1^{(1)}(s) e^{i{\mathcal L}_b (T_p+\tau')} \tilde H_{eb}(\tau)\right\}\nonumber \\
&&\hspace{.2in}-\int_0^\tau ds\ \tilde g_3^{(1)}(s,T_p,\tau') Tr_b\left\{\rho_b \tilde H_{eb}(s)\bar{H}_{eb}(\tau)\right\}\ . 
\een
The initial conditions for the above equations are $\tilde g_1^{(1)}(0)=|D_2\rangle \langle D_1|$ and  $\tilde g_3^{(1)}(0,T_p,\tau')=|D_4\rangle\langle D_3|g_1^{(1)}(\tau')$.

\subsection{Calculation of $\chi^{(2)}(\tau,T_p,\tau')$}

For the calculation of $\chi^{(2)}(\tau,T_p,\tau')$, 
we define 
\ben 
&&G_1^{(2)}(\tau')\equiv e^{-ih_{ex}\tau'/\hbar}\rho_b|D_1\rangle \langle D_2|e^{iH_b\tau'/\hbar}\ ,\\
&&G_2^{(2)}(T_p,\tau')\equiv e^{-ih_{ex} T_p/\hbar}G_1^{(3)}(\tau'')e^{ih_{ex} T_p/\hbar} \ ,\\
&&G_3^{(2)}(\tau,T_p,\tau')\equiv e^{-iH_b\tau/\hbar}|D_4\rangle\langle D_3|G_2^{(2)}(T_p,\tau')e^{ih_{ex}\tau/\hbar}\perd\nonumber \\
\een
The traces of these operators over the bath degrees of freedom are 
respectively defined as $g_1^{(2)}(\tau')$, $g_2^{(2)}(T_p,\tau')$, 
and $g_3^{(2)}(\tau,T_p,\tau')$.   
Then, 
\be
\chi^{(2)}(\tau,T_p,\tau')=Tr_e\{g_3^{(2)}(\tau,T_p,\tau')\}\perd
\ee
Explicit time evolution with respect to all of $\tau$, $T_p$, and $\tau'$ are necessary in the present case. Appendix B.2 provides exact QMEs for the following interaction picture system operators:
\ben
&&\tilde g_1^{(2)}(\tau')=e^{ih_e\tau'/\hbar}g_1^{(2)}(\tau') \ ,\\
&&\tilde g_2^{(2)}(T_p,\tau')=e^{ih_eT_p/\hbar}g_2(T_p,\tau')e^{-ih_eT_p/\hbar}\ , \\
&&\tilde g_3^{(2)}(\tau,T_p,\tau')=g_3^{(2)}(\tau,T_p,\tau')e^{-ih_e\tau/\hbar} \ .
\een  
Up to the 2nd order of $\tilde H_{eb}$, the formally exact QMEs, Eqs. (\ref{eq:g3p-2}), (\ref{eq:g22}), and (\ref{eq:g12}),  can be approximated as
\ben
&&\frac{d}{d\tau'} \tilde g_1^{(2)}(\tau')\nonumber \\
&&  \  \approx -\frac{1}{\hbar^2} \int_0^{\tau'} ds Tr_b\left\{\rho_b \tilde H_{eb}(\tau') \tilde H_{eb}(s) \right\} \tilde g_1^{(2)} (s) \ , \\
&&\frac{d}{dT_p} \tilde g_2^{(2)}(T_p,\tau')\approx \tilde I_2^{(2)}(T_p,\tau') \nonumber \\
&&\hspace{.2in} -\int_0^{T_p} ds Tr_b \left\{ \tilde {\mathcal L}_{eb}(T_p) \tilde {\mathcal L}_{eb}(s) \rho_b\right\} \tilde g_2^{(2)} (s,\tau')  \ ,
\een
where 
\ben
&&\tilde I_2^{(2)}(T_p,\tau')\approx -\frac{1}{\hbar}\int_0^{\tau'} ds\ Tr_b\left\{ \tilde {\mathcal L}_{eb}(T_p) e^{-i{\mathcal L}_b \tau'} \right .  \nonumber \\
&&\left . \hspace{1.in}\times e^{-ih_e \tau'/\hbar}\tilde H_{eb}(s) \rho_b \tilde g_1^{(2)}(s)\right\} , 
\een
and 
\ben
&&\frac{d}{d\tau} \tilde g_3^{(2)}(\tau, T_p,\tau')\approx \tilde I_3^{(2)}(\tau,T_p,\tau') \nonumber \\
&&\hspace{.2in} -\frac{1}{\hbar^2}\int_0^{\tau} ds  \tilde g_3^{(2)}(s,T_p,\tau') Tr_b \left\{ \tilde H_{eb}(s) \tilde H_{eb}(\tau) \rho_b\right\} \ , \nonumber \\
\een
with
\ben
&& \tilde I_3^{(2)}(\tau,T_p,\tau')\approx \frac{1}{\hbar^2}\int_0^{\tau'} ds\ Tr_b\left\{ \left (e^{-i{\mathcal L}_b T_p} e^{-i{\mathcal L}_e T_p}  \right . \right . \nonumber \\
&&\times \left . \left . e^{-i{\mathcal L}_b \tau'} e^{-ih_e \tau'/\hbar} \tilde H_{eb}(s) \rho_b \tilde g_1^{(2)} (s)\right ) \tilde H_{eb}(\tau) \right\} \nonumber \\
&&+\frac{1}{\hbar}\int_0^{T_p} ds\ Tr_b\left\{ \left (e^{-i{\mathcal L}_b T_p} e^{-i{\mathcal L}_e T_p}\tilde {\mathcal L}_{eb} (s) \tilde g_2^{(2)}(s,\tau') \rho_b \right ) \right  . \nonumber \\
&&\hspace{1.5in} \left . \times \tilde H_{eb}(\tau)\right\}\ .
\een
The initial conditions are such that $\tilde g_3^{(2)}(0,T_p,\tau')=|D_4\rangle\langle D_3|g_2(T_p,\tau')$,  $\tilde g_2^{(2)}(0,\tau')=g_1^{(2)}(\tau')$, and $g_1^{(2)}(0)=|D_1\rangle\langle D_2|$.

\subsection{Calculation of $\chi^{(3)}(\tau,T_p,\tau')$}

For the calculation of $\chi^{(3)}(\tau,T_p,\tau')$, 
we can define 
\ben 
&&G_1^{(3)}(\tau')\equiv e^{-iH_b\tau'/\hbar}\rho_b|D_2\rangle \langle D_1|e^{ih_{ex}\tau'/\hbar}\ ,\\
&&G_2^{(3)}(T_p,\tau')\equiv e^{-ih_{ex}T_p/\hbar}G_1^{(3)}(\tau')e^{ih_{ex}T_p/\hbar} \ ,\\
&&G_3^{(3)}(\tau,T_p,\tau')\equiv e^{-iH_b\tau/\hbar}|D_4\rangle\langle D_3|G_2^{(3)}(T_p,\tau')e^{ih_{ex}\tau/\hbar} \perd \nonumber \\
\een
The traces of these operators over the bath degrees of freedom are 
respectively defined as $g_1^{(3)}(\tau')$, $g_2^{(3)}(T_p,\tau')$, 
and $g_3^{(3)}(\tau,T_p,\tau')$.   
Then, 
\be
\chi^{(3)}(\tau,T_p,\tau')=Tr_e\{g_3^{(3)}(\tau,T_p,\tau')\}\perd
\ee
Three coupled equations are needed in this case as well.  Appendix B.3 provides exact QMEs for 
\ben 
&&\tilde g_1^{(3)}(\tau')=g_1^{(3)}(\tau')e^{-i h_e\tau'/\hbar}\ , \\
&&\tilde g_2^{(3)}(T_p,\tau')=e^{ih_eT_p/\hbar}g_2^{(3)}(T_p,\tau')e^{-ih_eT_p/\hbar}\ , \\
&&\tilde g_3^{(3)}(\tau,T_p,\tau')=g_3^{(3)}(\tau,T_p,\tau')e^{-ih_e\tau/\hbar}\ .
\een
Up to the second order of $\tilde H_{eb}$, the exact QMEs, Eqs. (\ref{eq:g3p-3}), (\ref{eq:g23}), and (\ref{eq:g13}),  can be approximated as
\ben
&& \frac{d}{d\tau'}\tilde g_1^{(3)}(\tau')\approx \nonumber \\
&&-\frac{1}{\hbar^2}\int_0^{\tau'} ds\  \tilde g_1^{(3)}(s) Tr_b\left\{\rho_b \tilde H_{eb}(s) \tilde H_{eb}(\tau' )\right\}\ , \nonumber \\  \\
&&\frac{d}{dT_p} \tilde g_2^{(3)}(T_p,\tau')\approx \tilde I_2^{(3)}(T_p,\tau') \nonumber \\
&&\hspace{.2in} -\int_0^{T_p} ds Tr_b \left\{ \tilde {\mathcal L}_{eb}(T_p) \tilde {\mathcal L}_{eb}(s) \rho_b\right\} \tilde g_2^{(3)} (s,\tau')\ , \nonumber  \\
\een
where
\ben
&&\tilde I_2^{(3)}(T_p,\tau')\approx -\frac{1}{\hbar}\int_0^{\tau'} ds\ Tr_b\left\{ \tilde {\mathcal L}_{eb}(T_p) e^{-i{\mathcal L}_b \tau'}e^{-ih_e \tau'/\hbar} \right .  \nonumber \\
&&\left . \hspace{1.5in}\times \rho_b \tilde g_1^{(3)}(s) \tilde H_{eb}(s)\right\}\ ,
\een
and
\ben
&&\frac{d}{d\tau} \tilde g_3^{(3)}(\tau, T_p,\tau')\approx \tilde I_3^{(3)}(\tau,T_p,\tau') \nonumber \\
&&\hspace{.2in} -\frac{1}{\hbar^2}\int_0^{\tau} ds  \tilde g_3^{(3)}(s,T_p,\tau') Tr_b \left\{ \tilde H_{eb}(s) \tilde H_{eb}(\tau) \rho_b\right\}\ ,  \nonumber \\
\een
with
\ben
&& \tilde I_3^{(3)}(\tau,T_p,\tau')\approx \frac{1}{\hbar^2}\int_0^{\tau'} ds\ Tr_b\left\{ \left (e^{-i{\mathcal L}_b T_p} e^{-i{\mathcal L}_e T_p}  \right . \right . \nonumber \\
&&\times  \left . e^{-i{\mathcal L}_b \tau'} e^{-ih_e \tau'/\hbar}\rho_b \tilde g_1^{(3)}(s)  \tilde H_{eb}(s) \tilde H_{eb}(\tau) \right\} \nonumber \\
&&+\frac{1}{\hbar}\int_0^{T_p} ds\ Tr_b\left\{ \left (e^{-i{\mathcal L}_b T_p} e^{-i{\mathcal L}_e T_p}\tilde {\mathcal L}_{eb} (s) \tilde g_2^{(3)}(s,\tau') \rho_b \right ) \right  . \nonumber \\
&&\hspace{1.5in} \left . \times \tilde H_{eb}(\tau)\right\} \ .
\een
The initial conditions for the above equations are $\tilde g_3^{(3)}(0,T_p,\tau')=|D_4\rangle\langle D_3|g_2^{(3)}(T_p,\tau')$, $\tilde g_2^{(3)}(0,\tau')=g_1^{(3)}(\tau')$, and $g_1(0)=|D_2\rangle\langle D_1|$.

\subsection{Calculation of $\chi^{(4)}(\tau,T_p,\tau')$}
For the calculation of $\chi^{(4)}(\tau,T_p,\tau')$,
we can define
\ben 
&&G_1^{(4)}(\tau')\equiv e^{-iH_b\tau'/\hbar}\rho_b|D_4\rangle \langle D_1|e^{ih_{ex}\tau'/\hbar}\\
&&G_2^{(4)}(T_p,\tau')\equiv e^{-iH_bT_p/\hbar}G_1^{(4)}(\tau')e^{iH_bT_p/\hbar} \\
&&G_3^{(4)}(\tau,T_p,\tau')\equiv e^{-iH_b\tau/\hbar}G_2^{(4)}(T_p,\tau')|D_2\rangle\langle D_3|e^{ih_{ex}\tau/\hbar} \nonumber \\
\perd
\een
The traces of these operators over the bath degrees of freedom are respectively denoted as  $g_1^{(4)}(\tau')$, $g_2^{(4)}(T_p,\tau')$, and $g_3^{(4)}(\tau,T_p,\tau')$.
Then, 
\be
\chi^{(4)}(\tau,T_p,\tau')=Tr_e\{g_3^{(4)}(\tau,T_p,\tau')\}\ .
\ee
As in the case of $\chi^{(1)}(\tau,T_p,\tau')$ explicit time evolutions are necessary only for $\tau$ and $\tau'$.   Appendix B.4 provides the exact QMEs for 
\ben
&&\tilde g_1^{(4)}(\tau')=g_1^{(4)}(\tau')e^{-i h_e \tau'/\hbar} \ , \\
&&\tilde g_3^{(4)}(\tau,T_p,\tau')=g_3^{(4)}(\tau,T_p,\tau')e^{-ih_e\tau/\hbar} \ .
\een
Within the 2nd order approximation, all the time evolution operators involving ${\mathcal Q}_L \tilde H_{eb}$ or $\tilde H_{eb} {\mathcal Q}_R$ can be disregarded in Eqs. (\ref{eq:g1p-4}) and (\ref{eq:g3p-4-final}).  Thus, we obtain the following approximations: 
\ben
&&\frac{d}{d\tau'}\tilde g_1^{(4)}(\tau')\approx \nonumber \\
&&-\frac{1}{\hbar^2}\int_0^{\tau'} ds\  \tilde g_1^{(4)}(s) Tr_b\left\{\rho_b \tilde H_{eb}(s) \tilde H_{eb}(\tau' )\right\}\ ,  \\
 &&\frac{d}{d\tau}\tilde g_3^{(4)}(\tau,T_p,\tau')= -\frac{1}{\hbar^2} \int_0^{\tau'} ds Tr_b\left\{ \left ( \tilde \rho_b g_1^{(4)}(s) \tilde H_{eb}(s) \right . \right . \nonumber \\
&&\hspace{.5in} \left.  \left . \times  e^{i H_e\tau'/\hbar} |D_2\rangle\langle D_3|  \right ) e^{i{\mathcal L}_b (T_p+\tau')}  \tilde H_{eb}(\tau)\right\} \nonumber \\
&&\hspace{.2in}-\int_0^\tau ds\ \tilde g_3^{(1)}(s,T_p,\tau') Tr_b\left\{\rho_b \tilde H_{eb}(s)\bar{H}_{eb}(\tau)\right\}\ .
\een
The initial conditions are such that $\tilde g_1^{(4)}(0)=|D_4\rangle \langle D_1|$ and $\tilde g_3^{(4)}(0,T_p,\tau')=g_1^{(4)}(\tau')|D_2\rangle \langle D_3|$.  These conclude the derivation of all the multistep QMEs necessary for the calculation of response functions.

\section{Conclusion}
This work provided a general formalism of four wave mixing spectroscopy of MCMM systems, and developed a new multistep QME approach for the calculation of third order response functions.  The consideration was limited to single exciton states only, but this was not due to fundamental limitation of the formalism but in order to present  the main idea in its simplest form. Thus, extension of the present work to include double exciton states is possible, which will be considered in the future.  In addition, explicit expressions for the response functions were derived for harmonic oscillator bath diagonally coupled to exciton states.   While this result is not new, its derivation based on polaron displacement operator is new.   The value of this derivation is more heuristic than practical, but it provides important insights  for developing approximations for more challenging cases  with off-diagonal exicton-bath coupling.

As was stated in the Introduction, the motivation of the present work was to develop an efficient and reliable computational methods that allow quantitative  modeling of modern 2DES spectroscopy.   
For an MCMM system where dephasing of exciton states due to diagonal exciton-bath couplings are the dominant mechanisms of line broadening, the results presented in Sec. III.B already serve that role.  The expressions for the response functions remain valid for any kind of spectral densities with or without correlations among different  site excitation or delocalized exciton states.  In addition, averaging of that expression over an ensemble of the disorder is possible at least numerically, which allows more quantitative assessment of homogeneous or inhomogeneous broadening mechanisms of detailed 2DES signals. 

Section IV amounts to the main result of the present work. The multistep QMEs can describe inter-exciton transitions due to bath-mediated coupling as well as radiation induced ones in a consistent manner while including all the effects of dephasing and relaxation mechanisms.  The complicated forms of the QMEs reflect the physical nature of the problem.    Multiple matter-radiation interactions alter the Hamiltonian governing the system during time intervals in-between, which are represented by different QMEs.   However, memory effects of the bath are sustained across the  matter-radiation interactions, which are taken care of by inhomogeneous terms.  

While the complete derivation of multistep QMEs was a significant step forward, much work still needs to be done to establish it as a general methodology.  As was indicated in the beginning of this section, generalization of this approach to include double exciton states is necessary.   Numerical tests for simple model systems are also important in order to understand what are the unique features that can be explained by the multistep QMEs.   Given that these objectives are accomplished, the formalism of the present work can serve as an attractive theoretical tool for quantitative analysis of various 2DES results for MCMM systems.

\acknowledgments
SJ is indebted to Prof. Kook Joe Shin for introducing him to the field of theoretical chemistry and for continuing encouragement and guidance.
This work was supported by the US National Science Foundation CAREER award (Grant No. CHE-0846899), the Office of Basic Energy Sciences, Department of Energy (Grant No. DE-SC0001393), and the Camille Dreyfus Teacher Scholar Award.

\appendix 

\section{Derivation of response functions for diagonal exciton-bath coupling} 
Consider the first response function $\chi^{(1)}(\tau,T_p,\tau')$ given by Eq. (\ref{eq:resp1}).   For the case where the exciton-bath coupling is diagonal in the exciton basis as shown in Eq. (\ref{eq:heb-m1}), $e^{-ih_{ex}\tau'/\hbar}$ and $e^{ih_{ex}\tau/\hbar}$ can be expanded in the basis of $|\varphi_j\rangle$'s.  The resulting expression is 
\ben
&&\chi^{(1)}(\tau,T_p,\tau')=\sum_{j}\sum_{j'}Tr_b\left\{e^{-iH_b(\tau+T_p)/\hbar} \langle D_2|\varphi_j\rangle \right . \nonumber \\
&&\hspace{.3in}\times e^{-i({\mathcal E}_j+\delta H_{bj}+H_b)\tau'/\hbar}\langle \varphi_j|D_1\rangle  \rho_b e^{iH_b(T_p+\tau')/\hbar}\nonumber \\
&&\hspace{.3in}\left .  \langle D_3|\varphi_{j'}\rangle e^{i({\mathcal E}_{j'}+\delta H_{bj'}+H_b)\tau/\hbar}\langle \varphi_{j'}|D_4\rangle \right\} \ ,\label{eq:chi1-ap1}  
\een
Let us introduce the following generator of polaron transformation for the exciton state $j$:
\be
S_j=-\sum_n g_{j,n} (b_n-b_n^\dagger)
\ee
Then, one can show that 
\be
e^{S_j}(H_b+\delta H_{bj})e^{-S_j}=H_b-\sum_n g_{j,n}^2 \hbar \omega_n
\ee
In Eq. (\ref{eq:chi1-ap1}), inserting $1=e^{-S_j}e^{S_j}$ before and after $e^{-i({\mathcal E}_j+\delta H_{bj}+H_b)\tau'/\hbar}$ and inserting $1=e^{-S_{j'}}e^{S_{j'}}$ before and after $e^{i({\mathcal E}_{j'}+\delta H_{bj'}+H_b)\tau/\hbar}$, we find the following expression:
\ben
&&\chi^{(1)}(\tau,T_p,\tau')=\sum_{j}\sum_{j'}D_{2,j}^*D_{1,j}D_{3,j'}^*D_{4,j'}\nonumber \\
&&\hspace{.3in}\times e^{-i\tilde {\mathcal E}_j\tau'/\hbar+i\tilde {\mathcal E}_{j'}\tau/\hbar}Tr_b\left\{e^{-S_{j'}(T_p+\tau')}\right . \nonumber \\
&&\hspace{.3in}\left. \times e^{S_{j'}(\tau+T_p+\tau')}e^{-S_j(\tau')}e^{S_j}\rho_b\right\} \ ,\label{eq:chi1-m1}  
\een
where 
\ben 
&&D_{\alpha,j}=\langle \varphi_j|D_\alpha\rangle \ , \label{eq:dal_j}\\
&&\tilde {\mathcal E}_j={\mathcal E}_j - \sum_n g_{j,n}^2 \hbar \omega_n \ , \label{eq:til_ej} \\
&&S_j(\tau)=-\sum_n g_{j,n}(b_n e^{-i\omega_n \tau}-b_n^\dagger e^{i\omega_n \tau}) \ . 
\een
In deriving Eq. (\ref{eq:chi1-m1}), the following identity has been used:
\be
e^{iH_b\tau /\hbar}e^{\pm S_j} e^{-iH_b\tau /\hbar} =e^{\pm S_j(\tau)} \ .
\ee
The product of four displacement operators with different time arguments in Eq. (\ref{eq:chi1-m1}) can be calculated using the fact that $e^Ae^B=e^{A+B}e^{[A,B]/2}$, an identity that holds between any operator $A$ and $B$ as long as $[A,B]$ commutes with $A$ and $B$.  In fact, the following general identity can be established: 
\begin{widetext}
\ben
&&Tr_b\left\{ e^{-S_{j'}(x)}e^{S_{j'}(x')} e^{-S_{j} (y)}e^{S_{j} (y)}\rho_b\right\} \nonumber \\
&&=e^{-\sum_n g_{j'n}^2 \left\{ \coth (\frac{\beta\hbar\omega_n}{2}) (1-\cos (\omega_n (x-x')))+i\sin (\omega_n (x-x'))\right\}} \nonumber \\
&&\times e^{-\sum_n g_{jn}^2 \left\{ \coth (\frac{\beta\hbar\omega_n}{2}) (1-\cos (\omega_n (y-y')))+i\sin (\omega_n (y-y'))\right\}} \nonumber \\
&&\times e^{-\sum_n g_{j'n}g_{jn}\coth(\frac{\beta \hbar \omega_n}{2}) \{\cos(\omega_n (x-y)) -\cos (\omega_n (x'-y))-\cos (\omega_n(x-y'))+\cos (\omega_n (x'-y'))\}} \nonumber \\
&&\times e^{i\sum_n g_{j'n}g_{jn}\{\sin(\omega_n (x-y)) -\sin (\omega_n (x'-y))-\sin (\omega_n(x-y'))+\sin (\omega_n (x'-y'))\}} \label{eq:disp-4-id}
\een
\end{widetext}
Application of the above identity to Eq. (\ref{eq:chi1-m1}) leads to Eq. (\ref{eq:chi1-m1-f}). 

For other response functions, similar manipulations lead to the following expressions:
\ben
&&\chi^{(2)}(\tau,T_p,\tau')=\sum_j\sum_{j'} D_{3,j}^*D_{1,j}D_{2,j'}^*D_{4,j'}\nonumber \\
&&\hspace{.3in}\times  e^{-i\tilde {\mathcal E}_j(T_p+\tau')/\hbar+i\tilde {\mathcal E}_{j'}(\tau+T_p)/\hbar}Tr_b\left\{e^{-S_{j'}(\tau')}\right .\nonumber \\
&&\hspace{.3in}\times \left . e^{S_{j'}(\tau+T_p+\tau')}e^{-S_j(T_p+\tau')}e^{S_j}\rho_b\right\}\comma \label{eq:chi2-m1}
\een
\ben
&&\chi^{(3)}(\tau,T_p,\tau')=\sum_j\sum_{j'} D_{3,j}^*D_{2,j}D_{1,j'}^*D_{4,j'}\nonumber \\
&&\hspace{.3in}\times e^{-i\tilde {\mathcal E}_j\tau'/\hbar+i\tilde {\mathcal E}_{j'}(\tau+T_p+\tau')/\hbar}Tr_b\left\{e^{-S_{j'}} \right. \nonumber \\
&&\hspace{.3in}\left .\times e^{S_{j'}(\tau+T_p+\tau')}e^{-S_j(T_p+\tau')}e^{S_j(\tau')}\rho_b\right\}\comma \label{eq:chi3-m1}
\een
\ben
&&\chi^{(4)}(\tau,T_p,\tau')=\sum_j\sum_{j'} D_{1,j}^*D_{2,j}D_{3,j'}^*D_{4,j'}\nonumber \\
&&\hspace{.3in}\times e^{i\tilde {\mathcal E}_j \tau'/\hbar+i\tilde {\mathcal E}_{j'} \tau/\hbar}Tr_b\left\{e^{-S_j} e^{S_j(\tau')}\right .\nonumber \\
&&\hspace{.3in} \left .\times e^{-S_{j'}(T_p+\tau')}e^{S_{j'}(\tau+T_p+\tau')}\rho_b\right\}\comma \label{eq:chi4-m1}
\een
Application of Eq. (\ref{eq:disp-4-id}) to the above expressions lead to Eqs. (\ref{eq:chi2-m1-f})-(\ref{eq:chi4-m1-f}). 

\section{Derivation of quantum master equations for general system-bath coupling}  
\subsection{Equations for $\chi^{(1)}(\tau,T_p,\tau')$}
First, define 
\ben
\tilde G_3^{(1)}(\tau,T_p,\tau')&\equiv& e^{iH_b\tau/\hbar}G_3^{(1)}(\tau,T_p,\tau')e^{-iH_b\tau/\hbar}e^{-ih_e\tau/\hbar}\nonumber \\
&=&e^{i{\mathcal L}_b\tau}G_3^{(1)}(\tau,T_p,\tau')e^{-ih_e\tau/\hbar}\ ,
\een
where ${\mathcal L}(\cdot)=[H_b,(\cdot)]/\hbar$.
Then,
\be
\frac{d}{d\tau}\tilde G_3^{(1)}(\tau,T_p,\tau')=\frac{i}{\hbar}\tilde G_3^{(1)}(\tau,T_p,\tau')\tilde H_{eb}(\tau)\comma
\ee
where
\ben
\tilde H_{eb}(\tau)&\equiv& e^{ih_e\tau/\hbar}e^{iH_b\tau/\hbar}H_{eb}e^{-iH_b\tau/\hbar}e^{-ih_e\tau/\hbar}\nonumber \\
&=&e^{i{\mathcal L}_b\tau}e^{i{\mathcal L}_e\tau}H_{eb}\perd
\een
We introduce a right-hand side projection operator such that $(\cdot){\mathcal P}_R=\rho_b Tr_b\{\cdot\}$ and ${\mathcal Q}_R=1-{\mathcal P}_R$.
Then, solving for $\tilde G_3^{(1)}(\tau,T_p,\tau'){\mathcal Q}_R$ and inserting
this into the time evolution equation of $\tilde G_3^{(1)}(\tau,T_p,\tau'){\mathcal P}_R$, we find that
\ben
&&\frac{d}{d\tau}\tilde g_3^{(1)}(\tau,T_p,\tau')=\nonumber \\
&& \frac{i}{\hbar}Tr_b\left\{\tilde G_3^{(1)}(0,T_p,\tau'){\mathcal Q}_R e_{(-)}^{\frac{i}{\hbar}\int_0^\tau ds \tilde H_{eb}(s){\mathcal Q}_R}\tilde H_{eb}(\tau)\right\}\nonumber \\
&&-\int_0^\tau ds\ \tilde g_3^{(1)}(s,T_p,\tau') \nonumber \\
&&\times Tr_b\left\{\rho_b \tilde H_{eb}(s){\mathcal Q}_R e_{(-)}^{\frac{i}{\hbar}\int_s^\tau ds'\tilde H_{eb}(s'){\mathcal Q}_R}\bar{H}_{eb}(\tau)\right\}\comma \label{eq:g3p-1}
\een

where
\ben
\tilde g_3^{(1)}(0,T_p,\tau')&=&|D_4\rangle\langle D_3|Tr_b\{G_1^{(1)}(\tau')\}\nonumber \\
&\equiv& |D_4\rangle\langle D_3|g_1^{(1)}(\tau')\comma \\
\tilde G_3^{(1)}(0,T_p,\tau'){\mathcal Q}_R&=&|D_4\rangle\langle D_3|\left (e^{-i{\mathcal L}_b T_p}G_1^{(1)}(\tau')\right ){\mathcal Q}_R \nonumber \\
&=&|D_4\rangle\langle D_3|e^{-i{\mathcal L}_b T_p}(G_1^{(1)}(\tau'){\mathcal Q}_R)\perd \nonumber \\ \label{eq:g3q-1}
\een
Thus, solution of Eq. (\ref{eq:g3p-1}) requires information on $g_1^{(1)}(\tau')$, which can be obtained from $G_1^{(1)}(\tau'){\mathcal P}_R$, and that on $G_1^{(1)}(\tau'){\mathcal Q}_R$.  The time evolutions for these can be obtained by employing a similar procedure.
Defining a similar interaction picture for $\tilde G_1^{(1)}(\tau')$ such that
\be
\tilde G_1^{(1)}(\tau')=e^{ih_e\tau'/\hbar}e^{i{\mathcal L}_b\tau'}G_1(\tau')\ . \label{eq:til_G11}
\ee
Then, employing a left hand projection operator ${\mathcal P}_L$ defined by ${\mathcal P}_L(\cdot)\equiv \rho_b Tr_b\left\{ (\cdot)\right\}$ and ${\mathcal Q}_L=1-{\mathcal P}_L$, and using the fact that ${\mathcal Q}_L \tilde G_1(0)=0$, we obtain the following equations:

\ben
&&{\mathcal Q}_L\tilde G_1^{(1)}(\tau') =-\frac{i}{\hbar}\int_0^{\tau'} ds\ e_{(+)}^{-\frac{i}{\hbar}\int_s^{\tau'} d s' {\mathcal Q}_L\tilde H_{eb}(s') }\nonumber \\
&&\hspace{1.5in} \times  {\mathcal Q}_L \tilde H_{eb}(s)\rho_b \tilde g_1^{(1)}(s)\label{eq:g1q}\\
&&\frac{d}{d\tau'}\tilde g_1^{(1)}(\tau')=-\frac{1}{\hbar^2}\int_0^{\tau'} ds\  Tr_b\left\{\rho_b \tilde H_{eb}(\tau') \right . \nonumber \\
&&\hspace{.5in} \left . \times  e_{(+)}^{-\frac{i}{\hbar}\int_s^{\tau'}ds' {\mathcal Q}_L \tilde H_{eb}(s')}{\mathcal Q}_L \tilde H_{eb}(s)\right\} \tilde g_1^{(1)}(s)\ .\nonumber \\ \label{eq:g1p}
\een

Combining Eqs. (\ref{eq:g3q-1}), (\ref{eq:til_G11}), and (\ref{eq:g1q}), we find that 
\ben
&&\tilde G_3^{(1)}(0,T_p,\tau'){\mathcal Q}_R=-\frac{i}{\hbar}|D_4\rangle\langle D_3| e^{-i{\mathcal L}_b (T_p+\tau')} e^{-i h_e \tau'/\hbar} \nonumber \\
&&\times \int_0^{\tau'} d s\ e_{(+)}^{-\frac{i}{\hbar}\int_s^{\tau'} d s' {\mathcal Q}_L\tilde H_{eb}(s') } \tilde H_{eb}(s)\rho_b \tilde g_1^{(1)}(s)\nonumber \ , \\
\een
where the fact that  $G_1^{(1)}(\tau'){\mathcal Q}_R={\mathcal Q}_L G_1^{(1)}(\tau')$ has been used. 
Inserting this into Eq. (\ref{eq:g3p-1}), we find that 
\ben
&&\frac{d}{d\tau}\tilde g_3^{(1)}(\tau,T_p,\tau')=- \frac{1}{\hbar^2}|D_4\rangle\langle D_3| e^{-i h_e \tau'/\hbar}   \nonumber \\
&&\times \int_0^{\tau'} ds Tr_b\left\{ \left ( e_{(+)}^{-\frac{i}{\hbar}\int_s^{\tau'} d s' {\mathcal Q}_L\tilde H_{eb}(s') }\tilde H_{eb}(s)\rho_b \tilde g_1^{(1)}(s)\right ) \right . \nonumber \\
&&\hspace{.5in} \times \left . e^{i{\mathcal L}_b (T_p+\tau')}  e_{(-)}^{\frac{i}{\hbar}\int_0^\tau ds \tilde H_{eb}(s){\mathcal Q}_R}\tilde H_{eb}(\tau)\right\}\nonumber \\
&&-\int_0^\tau ds\ \tilde g_3^{(1)}(s,T_p,\tau') \nonumber \\
&&\times Tr_b\left\{\rho_b \tilde H_{eb}(s){\mathcal Q}_R e_{(-)}^{\frac{i}{\hbar}\int_s^\tau ds'\tilde H_{eb}(s'){\mathcal Q}_R}\bar{H}_{eb}(\tau)\right\}\comma \label{eq:g3p-1-final}
\een

Equations (\ref{eq:g1q}), (\ref{eq:g1p}), and (\ref{eq:g3p-1-final}) form a closed set of equations  
that can be used to determine $\tilde g_3^{(1)}(\tau,T_p,\tau')$ starting from the initial condition, $g_1^{(1)}(0)=|D_2\rangle \langle D_1|$.  These equations are exact but impractical because calculation of the unprojected part is difficult to obtain.   Approximations of these equations up to the second order of $\tilde H_{eb}$ are provided in the main text.

\subsection{Equations for $\chi^{(2)}(\tau,T_p,\tau')$} 
Define 
\be
\tilde G_3^{(2)}(\tau, T_p,\tau') = e^{i{\mathcal L}_b \tau} G_3^{(2)}(\tau, T_p,\tau') e^{-i h_e \tau/\hbar} \ .
\ee
Then, $\tilde g_3(\tau,T_p,\tau')=Tr_b\{\tilde G_3^{(2)}(\tau, T_p,\tau')\}$. 
The equation governing the time evolution of $\tilde g_3^{(2)}(\tau,T_p,\tau')$ 
is the same as $\tilde g_3^{(1)}(\tau,T_p,\tau')$ except for the difference in the initial condition.
Thus, 
\ben
&&\frac{d}{d\tau}\tilde g_3^{(2)}(\tau,T_p,\tau')=\nonumber \\
&& \frac{i}{\hbar}Tr_b\left\{\tilde G_3^{(2)}(0,T_p,\tau'){\mathcal Q}_R e_{(-)}^{\frac{i}{\hbar}\int_0^\tau ds \tilde H_{eb}(s){\mathcal Q}_R}\tilde H_{eb}(\tau)\right\}\nonumber \\
&&-\int_0^\tau ds\ \tilde g_3^{(2)}(s,T_p,\tau') \nonumber \\
&&\times Tr_b\left\{\rho_b \tilde H_{eb}(s){\mathcal Q}_R e_{(-)}^{\frac{i}{\hbar}\int_s^\tau ds'\tilde H_{eb}(s'){\mathcal Q}_R}\bar{H}_{eb}(\tau)\right\}\comma \label{eq:g3p-2}
\een
where
\ben
\tilde g_3^{(2)}(0,T_p,\tau')&=&|D_4\rangle\langle D_3|Tr_b\left\{G_2^{(2)}(T_p,\tau')\right\}\nonumber \\
&=&|D_4\rangle\langle D_3|g_2^{(2)}(T_p,\tau')\\
\tilde G_3^{(2)}(0,T_p,\tau'){\mathcal Q}_R&=&|D_4\rangle\langle D_3|G_2^{(2)}(T_p,\tau'){\mathcal Q}_R \label{eq:g3_0q}
\een
Unlike the previous case, one has to solve explicitly the time evolution 
equations for $G_2^{(2)}(T_p,\tau')$.
Employing the left projection operator ${\mathcal P}_L(\cdot)=\rho_b Tr_b\left \{\cdot\right\}$ and ${\mathcal Q}_L=1-P_L$, this can be solved explicitly.  Thus, one can show that

\ben
&&{\mathcal Q}_L \tilde G_2^{(2)}(T_p,\tau')=e_{(+)}^{-i\int_0^{T_p} ds {\mathcal Q}_L \tilde {\mathcal L}_{eb}(s)} {\mathcal Q}_L \tilde G_2^{(2)}(0,\tau')\nonumber \\
&&-i\int_0^{T_p} ds\ e_{(+)}^{-i\int_s^{T_p} ds' {\mathcal Q}_L \tilde {\mathcal L}_{eb}(s')} {\mathcal Q}_L \tilde {\mathcal L}_{eb}(s)\rho_b \tilde g_2(s,\tau') \label{eq:qlg2_2}\\
&&\frac{d}{dT_p} \tilde g_2^{(2)}(T_p,\tau')=\nonumber \\
&&-iTr_b\left\{\tilde {\mathcal L}_{eb}(T_p)e_{(+)}^{-i\int_0^{T_p} ds {\mathcal Q}_L \tilde {\mathcal L}_{eb}(s)} {\mathcal Q}_L \tilde G_2(0,\tau')\right\}\nonumber \\
&&-\int_0^{T_p} ds\ Tr_b\left\{\tilde {\mathcal L}_{eb}(T_p)e_{(+)}^{-i\int_s^{T_p} ds' {\mathcal Q}_L \tilde {\mathcal L}_{eb}(s')} {\mathcal Q}_L \tilde {\mathcal L}_{eb}(s)\rho_b\right\}\nonumber \\
&&\hspace{1in}\times \tilde g_2^{(2)}(s,\tau')  \label{eq:g22}
\comma
\een 
where ${\mathcal Q}_L \tilde G_2^{(2)}(0,\tau')=G_1(\tau'){\mathcal Q}_R$ and $\tilde g_2^{(2)}(0,\tau')=g_1^{(1)}(\tau')$. The equations governing the time evolution of $G_1(\tau'){\mathcal Q}_R$ and 
$g_1(\tau')$ are the same as Eqs. (\ref{eq:g1q}) and (\ref{eq:g1p}), except
for the different initial condition, $G_1(0)=\rho_b|D_1\rangle\langle D_2|$.  
The resulting equations are as follows: 

\ben 
&&{\mathcal Q}_L \tilde G_1(\tau')=-\frac{i}{\hbar} \int_0^{\tau'} ds\ e_{(+)}^{-\frac{i}{\hbar} \int_0^{\tau'} ds' {\mathcal Q}_L \tilde H_{eb} (s')}\nonumber  \label{eq:qlg1}\\
&&\hspace{.5in} \times  {\mathcal Q}_L \tilde H_{eb} (s) \rho_b \tilde g_1^{(2)} (s)  \\
&&\frac{d}{d \tau'}\tilde g_1^{(2)}(\tau') =-\frac{1}{\hbar^2} \int_0^{\tau'} ds Tr_b\left \{ \rho_b \tilde H_{eb}(\tau')  \nonumber \right . \\ 
&&\hspace{.1in} \left . \times e_{(+)}^{-\frac{i}{\hbar}\int_s^{\tau'} ds' {\mathcal Q}_L \tilde H_{eb}(s')}{\mathcal Q}_L \tilde H_{eb}(s) \right \} \tilde g_1^{(2)}(s) \label{eq:g12}
\een
 
Using Eqs. (\ref{eq:g3_0q}) and (\ref{eq:qlg2_2}),  and the following identities: 
\ben 
&&G_2^{(2)}(T_p,\tau') {\mathcal Q}_R={\mathcal Q}_L G_2^{(2)}(T_p,\tau')  \nonumber \\
&&\hspace{.4in}={\mathcal Q}_L e^{-i {\mathcal L}_b T_p} e^{-i {\mathcal L}_e T_p} \tilde G_2^{(2)}(T_p,\tau') \nonumber \\
&&\hspace{.4in}=e^{-i {\mathcal L}_b T_p} e^{-i {\mathcal L}_e T_p} {\mathcal Q}_L \tilde G_2^{(2)}(T_p,\tau') \\
&&{\mathcal Q}_L \tilde G^{(2)}(0,\tau')={\mathcal Q}_L \tilde G_1^{(2)} (\tau') \nonumber \\
&&\hspace{.4in}={\mathcal Q}_L e^{-i{\mathcal L}_b \tau'} e^{-ih_e \tau'/\hbar} \tilde G_1^{(2)} (\tau') \nonumber \\
&&\hspace{.4in}=e^{-i {\mathcal L}_b \tau'} e^{-i h_e \tau'/\hbar} {\mathcal Q}_L \tilde G_1^{(2)}(\tau')\ , 
\een
the inhomogeneous term in Eq. (\ref{eq:g3p-2}) can be expressed as
\ben
&& \tilde I_3^{(2)}(\tau,T_p,\tau')\nonumber \\
&&\equiv \frac{i}{\hbar}Tr_b\left\{\tilde G_3^{(2)}(0,T_p,\tau'){\mathcal Q}_R e_{(-)}^{\frac{i}{\hbar}\int_0^\tau ds \tilde H_{eb}(s){\mathcal Q}_R}\tilde H_{eb}(\tau)\right\}\nonumber \\
&&=\frac{i}{\hbar} Tr_b \left\{ \left ( e^{-i{\mathcal L}_b T_p} e^{-i{\mathcal L}_e T_p} e_{(+)}^{-i \int_0^{T_p} ds\ {\mathcal Q}_L \tilde {\mathcal L}_{eb}(s)} e^{-i{\mathcal L}_b \tau'}  \nonumber \right .\right .\\
&& \left. \left . \hspace{.2in}\times e^{-i h_e \tau'/\hbar}  {\mathcal Q}_L \tilde G_1^{(2)} (\tau')\right ) e_{(-)}^{\frac{i}{\hbar}\int_0^\tau ds\ \tilde H_{eb} (s) {\mathcal Q}_R }\tilde H_{eb}(\tau) \right\} \nonumber \\
&&+\frac{1}{\hbar} \int_0^{T_p} Tr_b ds \left\{ \left ( e^{-i{\mathcal L}_b T_p} e^{-i{\mathcal L}_e T_p} e_{(+)}^{-i \int_s^{T_p} ds'\ {\mathcal Q}_L \tilde {\mathcal L}_{eb}(s')}  \nonumber \right .\right .\\
&& \left. \left . \hspace{.2in}\times {\mathcal Q}_L \tilde {\mathcal L}_{eb}(s) \tilde g_2^{(2)} (s,\tau')\rho_b\right ) e_{(-)}^{\frac{i}{\hbar}\int_0^\tau ds\ \tilde H_{eb} (s) {\mathcal Q}_R }\tilde H_{eb}(\tau) \right\}\nonumber \\
\een 
Inserting Eq. (\ref{eq:qlg1}) into the first of the above equation, we obtain the following final expression:
\ben
&& \tilde I_3^{(2)}(\tau,T_p,\tau')\nonumber \\
&&=\frac{1}{\hbar^2} \int_0^{\tau'} ds\ Tr_b \left\{ \left ( e^{-i{\mathcal L}_b T_p} e^{-i{\mathcal L}_e T_p} e_{(+)}^{-i \int_0^{T_p} ds\ {\mathcal Q}_L \tilde {\mathcal L}_{eb}(s)}    \nonumber \right .\right .\\
&& \hspace{.2in}\times  e^{-i{\mathcal L}_b \tau'} e^{-i h_e \tau'/\hbar}  e_{(+)}^{-\frac{i}{\hbar} \int_0^{\tau'} ds' {\mathcal Q}_L \tilde H_{eb} (s') } \nonumber \\
&&\hspace{.2in}\left . \left . {\mathcal Q}_L  \tilde H_{eb} (s) \rho_b \tilde g_1^{(2)} (s) \right ) \times e_{(-)}^{\frac{i}{\hbar}\int_0^\tau ds\ \tilde H_{eb} (s) {\mathcal Q}_R }\tilde H_{eb}(\tau) \right\} \nonumber \\
&&+\frac{1}{\hbar} \int_0^{T_p} ds Tr_b \left\{ \left ( e^{-i{\mathcal L}_b T_p} e^{-i{\mathcal L}_e T_p} e_{(+)}^{-i \int_s^{T_p} ds'\ {\mathcal Q}_L \tilde {\mathcal L}_{eb}(s')}  \nonumber \right .\right .\\
&& \left. \left . \hspace{.2in}\times {\mathcal Q}_L \tilde {\mathcal L}_{eb}(s) \tilde g_2^{(2)} (s,\tau')\rho_b\right ) e_{(-)}^{\frac{i}{\hbar}\int_0^\tau ds\ \tilde H_{eb} (s) {\mathcal Q}_R }\tilde H_{eb}(\tau) \right\}\nonumber \\
\een
Similarly, the inhomogeneous term of Eq. (\ref{eq:g22}) can be shown to be 
\ben
&&\tilde I_2^{(2)}(T_p,\tau') \nonumber \\
&&\equiv -iTr_b\left\{\tilde {\mathcal L}_{eb}(T_p)e_{(+)}^{-i\int_0^{T_p} ds {\mathcal Q}_L \tilde {\mathcal L}_{eb}(s)} {\mathcal Q}_L \tilde G_2(0,\tau')\right\} \nonumber \\
&&=-\frac{1}{\hbar}\int_0^{\tau'}ds Tr_b\left\{ \tilde {\mathcal L}_{eb} (T_p) e_{(+)}^{-i\int_0^{T_p} ds\ {\mathcal Q}_L \tilde {\mathcal L}_{eb} (s) } e^{-i{\mathcal L}_{eb}\tau'}  \nonumber \right . \\
&&\left .\times  e^{-ih_e \tau'/\hbar} e_{(+)}^{-\frac{i}{\hbar} \int_0^{\tau'} ds'\ {\mathcal Q}_L \tilde H_{eb} (s')}{\mathcal Q}_L \tilde H_{eb} (s) \rho_b \tilde g_1^{(2)} (s) \right\}\nonumber \\
\een

\subsection{Equations for $\chi^{(3)}(\tau,T_p,\tau')$} 
Define 
\be
\tilde G_3^{(3)}(\tau, T_p,\tau') = e^{i{\mathcal L}_b \tau} G_3^{(3)}(\tau, T_p,\tau') e^{-i h_e \tau/\hbar} \ .
\ee
Then, $\tilde g_3(\tau,T_p,\tau')=Tr_b\{\tilde G_3^{(3)}(\tau, T_p,\tau')\}$. 
The equation governing the time evolution of $\tilde g_3^{(3)}(\tau,T_p,\tau')$ 
is the same as $\tilde g_3^{(2)}(\tau,T_p,\tau')$ except for the difference in the initial condition.
Thus, 
\ben
&&\frac{d}{d\tau}\tilde g_3^{(3)}(\tau,T_p,\tau')=\nonumber \\
&& \frac{i}{\hbar}Tr_b\left\{\tilde G_3^{(3)}(0,T_p,\tau'){\mathcal Q}_R e_{(-)}^{\frac{i}{\hbar}\int_0^\tau ds \tilde H_{eb}(s){\mathcal Q}_R}\tilde H_{eb}(\tau)\right\}\nonumber \\
&&-\int_0^\tau ds\ \tilde g_3^{(3)}(s,T_p,\tau') \nonumber \\
&&\times Tr_b\left\{\rho_b \tilde H_{eb}(s){\mathcal Q}_R e_{(-)}^{\frac{i}{\hbar}\int_s^\tau ds'\tilde H_{eb}(s'){\mathcal Q}_R}\bar{H}_{eb}(\tau)\right\}\comma \label{eq:g3p-3}
\een
where
\ben
\tilde g_3^{(3)}(0,T_p,\tau')&=&|D_4\rangle\langle D_3|Tr_b\left\{G_2^{(3)}(T_p,\tau')\right\}\nonumber \\
&=&|D_4\rangle\langle D_3|g_2^{(3)}(T_p,\tau')\\
\tilde G_3^{(3)}(0,T_p,\tau'){\mathcal Q}_R&=&|D_4\rangle\langle D_3|G_2^{(3)}(T_p,\tau'){\mathcal Q}_R \label{eq:g3_0q}
\een
The time evolution 
equations for $G_2^{(3)}(T_p,\tau')$ are the same as $G_2^{(3)}(T_p,\tau')$.
Thus, one can show that
\ben
&&{\mathcal Q}_L \tilde G_2^{(3)}(T_p,\tau')=e_{(+)}^{-i\int_0^{T_p} ds {\mathcal Q}_L \tilde {\mathcal L}_{eb}(s)} {\mathcal Q}_L \tilde G_2^{(3)}(0,\tau')\nonumber \\
&&-i\int_0^{T_p} ds\ e_{(+)}^{-i\int_s^{T_p} ds' {\mathcal Q}_L \tilde {\mathcal L}_{eb}(s')} {\mathcal Q}_L \tilde {\mathcal L}_{eb}(s)\rho_b \tilde g_2^{(3)}(s,\tau') \label{eq:qlg2_3}\\
&&\frac{d}{dT_p} \tilde g_2^{(3)}(T_p,\tau')=\nonumber \\
&&-iTr_b\left\{\tilde {\mathcal L}_{eb}(T_p)e_{(+)}^{-i\int_0^{T_p} ds {\mathcal Q}_L \tilde {\mathcal L}_{eb}(s)} {\mathcal Q}_L \tilde G_2^{(3)}(0,\tau')\right\}\nonumber \\
&&-\int_0^{T_p} ds\ Tr_b\left\{\tilde {\mathcal L}_{eb}(T_p)e_{(+)}^{-i\int_s^{T_p} ds' {\mathcal Q}_L \tilde {\mathcal L}_{eb}(s')} {\mathcal Q}_L \tilde {\mathcal L}_{eb}(s)\rho_b\right\}\nonumber \\
&&\hspace{1in}\times \tilde g_2^{(3)}(s,\tau')  \label{eq:g23}
\comma
\een 
where ${\mathcal Q}_L \tilde G_2^{(3)}(0,\tau')=G_1^{(3)}(\tau'){\mathcal Q}_R$ and $\tilde g_2^{(3)}(0,\tau')=g_1^{(3)}(\tau')$. The equations governing the time evolution of $G_1^{(3)}(\tau'){\mathcal Q}_R$ and  $g_1^{(3)}(\tau')$ are derived below.

Define a similar interaction picture for $\tilde G_1^{(3)}(\tau')$ such that
\be
\tilde G_1^{(3)}(\tau')=e^{i{\mathcal L}_b\tau'}G_1^{(3)}(\tau')e^{-ih_e\tau'/\hbar}\ . \label{eq:til_G13}
\ee
Then,  we obtain the following equations:

\ben
&&\tilde G_1^{(3)}(\tau') {\mathcal Q}_R=\frac{i}{\hbar}\int_0^{\tau'} ds\ \tilde g_1^{(3)}(s) \tilde H_{eb}(s){\mathcal Q}_R \nonumber \\
&&\hspace{1.5in} \times e_{(-)}^{\frac{i}{\hbar}\int_s^{\tau'} d s' \ \tilde H_{eb}(s') {\mathcal Q}_R}\ , \label{eq:g1q-3}\\
&&\frac{d}{d\tau'}\tilde g_1^{(3)}(\tau')=-\frac{1}{\hbar^2}\int_0^{\tau'} ds\  \tilde g_1^{(3)}(s) Tr_b\left\{\rho_b \tilde H_{eb}(s) {\mathcal Q}_R  \right . \nonumber \\
&&\hspace{.5in} \left . \times  e_{(-)}^{\frac{i}{\hbar}\int_s^{\tau'}ds'  \tilde H_{eb}(s'){\mathcal Q}_R}\tilde H_{eb}(\tau') \right\} \ .\nonumber \\ \label{eq:g13}
\een 
The initial condition is such that $G_1^{(3)}(0)=\rho_b|D_2\rangle\langle D_1|$.  

Using Eqs. (\ref{eq:g3_0q}) and (\ref{eq:qlg2_2}),  and the following identity
\ben 
&&{\mathcal Q}_L \tilde G_2^{(3)}(0,\tau')={\mathcal Q}_L \tilde G_1^{(3)} (\tau') = \tilde G_1^{(3)} (\tau') {\mathcal Q}_R \nonumber \\
&&\hspace{.4in}=e^{-i {\mathcal L}_b \tau'}  \tilde G_1^{(3)}(\tau')  {\mathcal Q}_R e^{i h_e \tau'/\hbar}\ , 
\een
the inhomogeneous term in Eq. (\ref{eq:g3p-2}) can be expressed as
\ben
&& \tilde I_3^{(3)}(\tau,T_p,\tau')\nonumber \\
&&\equiv \frac{i}{\hbar}Tr_b\left\{\tilde G_3^{(2)}(0,T_p,\tau'){\mathcal Q}_R e_{(-)}^{\frac{i}{\hbar}\int_0^\tau ds \tilde H_{eb}(s){\mathcal Q}_R}\tilde H_{eb}(\tau)\right\}\nonumber \\
&&=\frac{1}{\hbar} \int_0^{\tau'} ds\ Tr_b \left\{ \left ( e^{-i{\mathcal L}_b T_p} e^{-i{\mathcal L}_e T_p} e_{(+)}^{-i \int_0^{T_p} ds\ {\mathcal Q}_L \tilde {\mathcal L}_{eb}(s)}    \nonumber \right .\right .\\
&& \hspace{.2in}\times   e^{-i{\mathcal L}_b \tau'} e^{-i h_e \tau'/\hbar} \tilde g_1^{(3)}(s) \tilde H_{eb}(s){\mathcal Q}_R      \nonumber  \\
&&\hspace{.2in}\left . \left .\times e_{(-)}^{\frac{i}{\hbar}\int_s^{\tau'} d s' \ \tilde H_{eb}(s')  {\mathcal Q}_R} \right ) e_{(-)}^{\frac{i}{\hbar}\int_0^\tau ds\ \tilde H_{eb} (s) {\mathcal Q}_R }\tilde H_{eb}(\tau) \right\} \nonumber \\
&&+\frac{1}{\hbar} \int_0^{T_p} Tr_b ds \left\{ \left ( e^{-i{\mathcal L}_b T_p} e^{-i{\mathcal L}_e T_p} e_{(+)}^{-i \int_s^{T_p} ds'\ {\mathcal Q}_L \tilde {\mathcal L}_{eb}(s')}  \nonumber \right .\right .\\
&& \left. \left . \hspace{.2in}\times {\mathcal Q}_L \tilde {\mathcal L}_{eb}(s) \tilde g_2^{(2)} (s,\tau')\rho_b\right ) e_{(-)}^{\frac{i}{\hbar}\int_0^\tau ds\ \tilde H_{eb} (s) {\mathcal Q}_R }\tilde H_{eb}(\tau) \right\}\nonumber \\
\een
Similarly, the inhomogeneous term of Eq. (\ref{eq:g22}) can be shown to be 
\ben
&&\tilde I_2^{(3)}(T_p,\tau') \nonumber \\
&&\equiv -iTr_b\left\{\tilde {\mathcal L}_{eb}(T_p)e_{(+)}^{-i\int_0^{T_p} ds {\mathcal Q}_L \tilde {\mathcal L}_{eb}(s)} {\mathcal Q}_L \tilde G_2(0,\tau')\right\} \nonumber \\
&&=\frac{1}{\hbar}\int_0^{\tau'}ds Tr_b\left\{ \tilde {\mathcal L}_{eb} (T_p) e_{(+)}^{-i\int_0^{T_p} ds\ {\mathcal Q}_L \tilde {\mathcal L}_{eb} (s) } e^{-i{\mathcal L}_{eb}\tau'}  \nonumber \right . \\
&&\times  e^{-ih_e \tau'/\hbar} \left ( \tilde g_1^{(3)}(s) \tilde H_{eb}(s){\mathcal Q}_R   \right .    \nonumber  \\
&&\hspace{.2in}\left . \left .\times e_{(-)}^{\frac{i}{\hbar}\int_s^{\tau'} d s' \ \tilde H_{eb}(s')  {\mathcal Q}_R}  \right ) \right\}\nonumber \\
\een

\subsection{Equations for $\chi^{(4)}(\tau,T_p,\tau')$}
The equations for this can be derived in a similar manner as $\chi^{(4)}(\tau,T_p,\tau')$.
First, define 
\ben
\tilde G_3^{(4)}(\tau,T_p,\tau')&\equiv& e^{iH_b\tau/\hbar}G_3^{(4)}(\tau,T_p,\tau')e^{-iH_b\tau/\hbar}e^{-ih_e\tau/\hbar}\nonumber \\
&=&e^{i{\mathcal L}_b\tau}G_3^{(4)}(\tau,T_p,\tau')e^{-ih_e\tau/\hbar}\ ,
\een
Then,
\ben
&&\frac{d}{d\tau}\tilde g_3^{(4)}(\tau,T_p,\tau')=\nonumber \\
&& \frac{i}{\hbar}Tr_b\left\{\tilde G_3^{(4)}(0,T_p,\tau'){\mathcal Q}_R e_{(-)}^{\frac{i}{\hbar}\int_0^\tau ds \tilde H_{eb}(s){\mathcal Q}_R}\tilde H_{eb}(\tau)\right\}\nonumber \\
&&-\int_0^\tau ds\ \tilde g_3^{(4)}(s,T_p,\tau') \nonumber \\
&&\times Tr_b\left\{\rho_b \tilde H_{eb}(s){\mathcal Q}_R e_{(-)}^{\frac{i}{\hbar}\int_s^\tau ds'\tilde H_{eb}(s'){\mathcal Q}_R}\bar{H}_{eb}(\tau)\right\}\comma \label{eq:g3p-4}
\een

where
\ben
\tilde g_3^{(4)}(0,T_p,\tau')&=&Tr_b\{G_1^{(4)}(\tau')\}|D_2\rangle\langle D_3|\nonumber \\
&\equiv& g_1^{(4)}(\tau')|D_2\rangle\langle D_3|\comma \\
\tilde G_3^{(4)}(0,T_p,\tau'){\mathcal Q}_R&=&\left (e^{-i{\mathcal L}_b T_p}G_1^{(4)}(\tau')\right )|D_2\rangle\langle D_3|{\mathcal Q}_R \nonumber \\
&=&e^{-i{\mathcal L}_b T_p}(G_1^{(4)}(\tau'){\mathcal Q}_R)|D_2\rangle\langle D_3|\perd \nonumber \\ \label{eq:g3q-4}
\een
Thus, solution of Eq. (\ref{eq:g3p-1}) requires information on $g_1^{(4)}(\tau')$, which can be obtained from $G_1^{(4)}(\tau'){\mathcal P}_R$, and that on $G_1^{(4)}(\tau'){\mathcal Q}_R$.  The time evolutions for these can be obtained by employing a similar procedure.
Defining a similar interaction picture for $\tilde G_1^{(4)}(\tau')$ such that
\be
\tilde G_1^{(4)}(\tau')=e^{i{\mathcal L}_b\tau'}G_1^{(4)}(\tau')e^{-ih_e\tau'/\hbar}\ . \label{eq:til_G14}
\ee
Then, employing a lefthand projection operator ${\mathcal P}_L$ defined by ${\mathcal P}_L(\cdot)\equiv \rho_b Tr_b\left\{ (\cdot)\right\}$ and ${\mathcal Q}_L=1-{\mathcal P}_L$, and using the fact that ${\mathcal Q}_L \tilde G_1(0)=0$, we obtain the following equations:

\ben
&&\tilde G_1^{(4)}(\tau') {\mathcal Q}_R=\frac{i}{\hbar}\int_0^{\tau'} ds\ \tilde g_1^{(4)}(s) \tilde H_{eb}(s){\mathcal Q}_R \nonumber \\
&&\hspace{1.5in} \times e_{(-)}^{\frac{i}{\hbar}\int_s^{\tau'} d s' \ \tilde H_{eb}(s') {\mathcal Q}_R}\nonumber \label{eq:g1q-4}\\
&&\frac{d}{d\tau'}\tilde g_1^{(4)}(\tau')=-\frac{1}{\hbar^2}\int_0^{\tau'} ds\  \tilde g_1^{(4)}(s) Tr_b\left\{\rho_b \tilde H_{eb}(s) {\mathcal Q}_R  \right . \nonumber \\
&&\hspace{.5in} \left . \times  e_{(-)}^{\frac{i}{\hbar}\int_s^{\tau'}ds'  \tilde H_{eb}(s'){\mathcal Q}_R}\tilde H_{eb}(\tau') \right\} \ .\nonumber \\ \label{eq:g1p-4}
\een

Combining Eqs. (\ref{eq:g3q-1}), (\ref{eq:til_G11}), and (\ref{eq:g1q}), we find that 
\ben
&&\tilde G_3^{(4)}(0,T_p,\tau'){\mathcal Q}_R=\frac{i}{\hbar} e^{-i{\mathcal L}_b (T_p+\tau')} \int_0^{\tau'} d s\  \tilde g_1^{(4)}(s) \tilde H_{eb}(s) \nonumber \\
&&\hspace{.4in} \times  e_{(-)}^{\frac{i}{\hbar}\int_s^{\tau'} d s' \ \tilde H_{eb}(s') {\mathcal Q}_R} e^{i H_e\tau'/\hbar} |D_2\rangle\langle D_3|\ . 
\een
 Inserting this into Eq. (\ref{eq:g3p-4}), we find that 
\ben
&&\frac{d}{d\tau}\tilde g_3^{(4)}(\tau,T_p,\tau')=- \frac{1}{\hbar^2} \int_0^{\tau'} ds Tr_b\left\{ \left ( \rho_b g_1^{(4)}(s) \tilde H_{eb}(s) \right . \right . \nonumber \\
&&\hspace{.5in} \left. \times e_{(-)}^{\frac{i}{\hbar}\int_s^{\tau'} d s' \ \tilde H_{eb}(s') {\mathcal Q}_R} e^{i H_e\tau'/\hbar} |D_2\rangle\langle D_3|  \right ) \nonumber \\
&&\hspace{.5in} \times \left . e^{i{\mathcal L}_b (T_p+\tau')}  e_{(-)}^{\frac{i}{\hbar}\int_0^\tau ds \tilde H_{eb}(s){\mathcal Q}_R}\tilde H_{eb}(\tau)\right\}\nonumber \\
&&-\int_0^\tau ds\ \tilde g_3^{(1)}(s,T_p,\tau') \nonumber \\
&&\times Tr_b\left\{\rho_b \tilde H_{eb}(s){\mathcal Q}_R e_{(-)}^{\frac{i}{\hbar}\int_s^\tau ds'\tilde H_{eb}(s'){\mathcal Q}_R}\bar{H}_{eb}(\tau)\right\}\ . \label{eq:g3p-4-final}
\een

\end{document}